\documentclass[journal]{IEEEtran}
\IEEEoverridecommandlockouts
\UseRawInputEncoding
\usepackage{cite}
\usepackage{amsmath,amssymb,amsfonts}
\usepackage{algorithmic}
\usepackage[linesnumbered,ruled,vlined]{algorithm2e}
\usepackage{graphicx}
\usepackage{textcomp}
\usepackage{xcolor}
\usepackage{subfigure}
\def\BibTeX{{\rm B\kern-.05em{\sc i\kern-.025em b}\kern-.08em
    T\kern-.1667em\lower.7ex\hbox{E}\kern-.125emX}}

\begin{document}





\title{DRL Optimization Trajectory Generation via Wireless Network Intent-Guided Diffusion Models for Optimizing Resource Allocation}

\author{Junjie~Wu,
    Xuming~Fang,~\IEEEmembership{Senior Member,~IEEE},        Dusit~Niyato,~\IEEEmembership{Fellow,~IEEE}, \\ Jiacheng~Wang,
        Jingyu~Wang

\thanks{Junjie Wu, Xuming Fang, and Jingyu Wang are with Key Lab of Info Coding \& Transmission, Southwest Jiaotong University, Chengdu 610031, China. (E-mails:junjie\_wu@my.swjtu.edu.cn; xmfang@swjtu.edu.cn; wangmr930@my.swjtu.edu.cn).

Dusit Niyato and Jiacheng Wang are with the School
of Computer Science and Engineering, Nanyang Technological University,
Singapore 639798 (e-mail: dniyato@ntu.edu.sg; jiacheng.wang@ntu.edu.sg).

This work was supported in part by the NSFC under Grant 62071393, and Fundamental Research Funds for the Central Universities under Grant 2682023ZTPY058.
}}



\maketitle

\begin{abstract}

With the rapid advancements in wireless communication fields, including low-altitude economies, 6G, and Wi-Fi, the scale of wireless networks continues to expand, accompanied by increasing service quality demands. Traditional deep reinforcement learning (DRL)-based optimization models can improve network performance by solving non-convex optimization problems intelligently. However, they heavily rely on online deployment and often require extensive initial training. Online DRL optimization models typically make accurate decisions based on current channel state distributions. When these distributions change, their generalization capability diminishes, which hinders the responsiveness essential for real-time and high-reliability wireless communication networks. Furthermore, different users have varying quality of service (QoS) requirements across diverse scenarios, and conventional online DRL methods struggle to accommodate this variability. Consequently, exploring flexible and customized AI strategies is critical. We propose a wireless network intent (WNI)-guided trajectory generation model based on a generative diffusion model (GDM). This model can be generated and fine-tuned in real time to achieve the objective and meet the constraints of target intent networks, significantly reducing state information exposure during wireless communication. Moreover, The WNI-guided optimization trajectory generation can be customized to address differentiated QoS requirements, enhancing the overall quality of communication in future intelligent networks. Extensive simulation results demonstrate that our approach achieves greater stability in spectral efficiency variations and outperforms traditional DRL optimization models in dynamic communication systems.

\end{abstract}


\section{Introduction}

Real-time and high-reliability wireless communication network technologies have gained widespread application in emergency communications \cite{add1}, unmanned aerial vehicles (UAVs) networks \cite{add2}, and environmental monitoring \cite{add3} in the low-altitude economy, as illustrated in Fig. 1. These applications aim to improve channel resource allocation efficiency and improve user service quality.  As the scale of wireless communication network deployments expands and related applications continue to evolve, wireless communication environments are becoming increasingly complex. Achieving objectives \cite{add4,add5} such as signal coverage optimization, flight trajectory estimation, and interference minimization among UAVs across various application scenarios presents significant challenges. These challenges stem from the dynamic and unpredictable nature of the operating environments and the need for efficient channel resource allocation. Furthermore, these objectives are inherently interdependent and must be jointly optimized under resource-constrained conditions to ensure the stable operation of the overall system. For example, the channel quality between UAVs and base stations (BSs), or among UAVs following different flight paths, can vary significantly. Therefore, UAV trajectory planning must account for channel conditions to ensure that the UAVs remain within areas of strong signal strength. 


Similarly, signal coverage requirements vary depending on environmental factors in communication scenarios involving 6G and next-generation Wi-Fi networks. Obstacles, geographical features, and user locations can impact wireless signal strength. As a result, channel resource allocation algorithms must adapt to these dynamic factors to maintain stable, high-quality communication services. 
In such cases, it is crucial to develop channel allocation algorithms tailored to specific objectives across different scenarios. These algorithms should enable the system to dynamically allocate resources based on the relative positions and mission priorities of the communication tasks. Thus, optimization objectives—such as signal coverage, trajectory planning, and interference management—are closely intertwined with channel resource allocation. Addressing these challenges requires real-time environmental sensing and adaptive resource allocation to meet the diverse demands of users across varying scenarios.

\begin{figure*}[]
  \begin{center}
  
  \subfigure[Emergency Communications]{
    \scalebox{0.22}[0.22]{\includegraphics{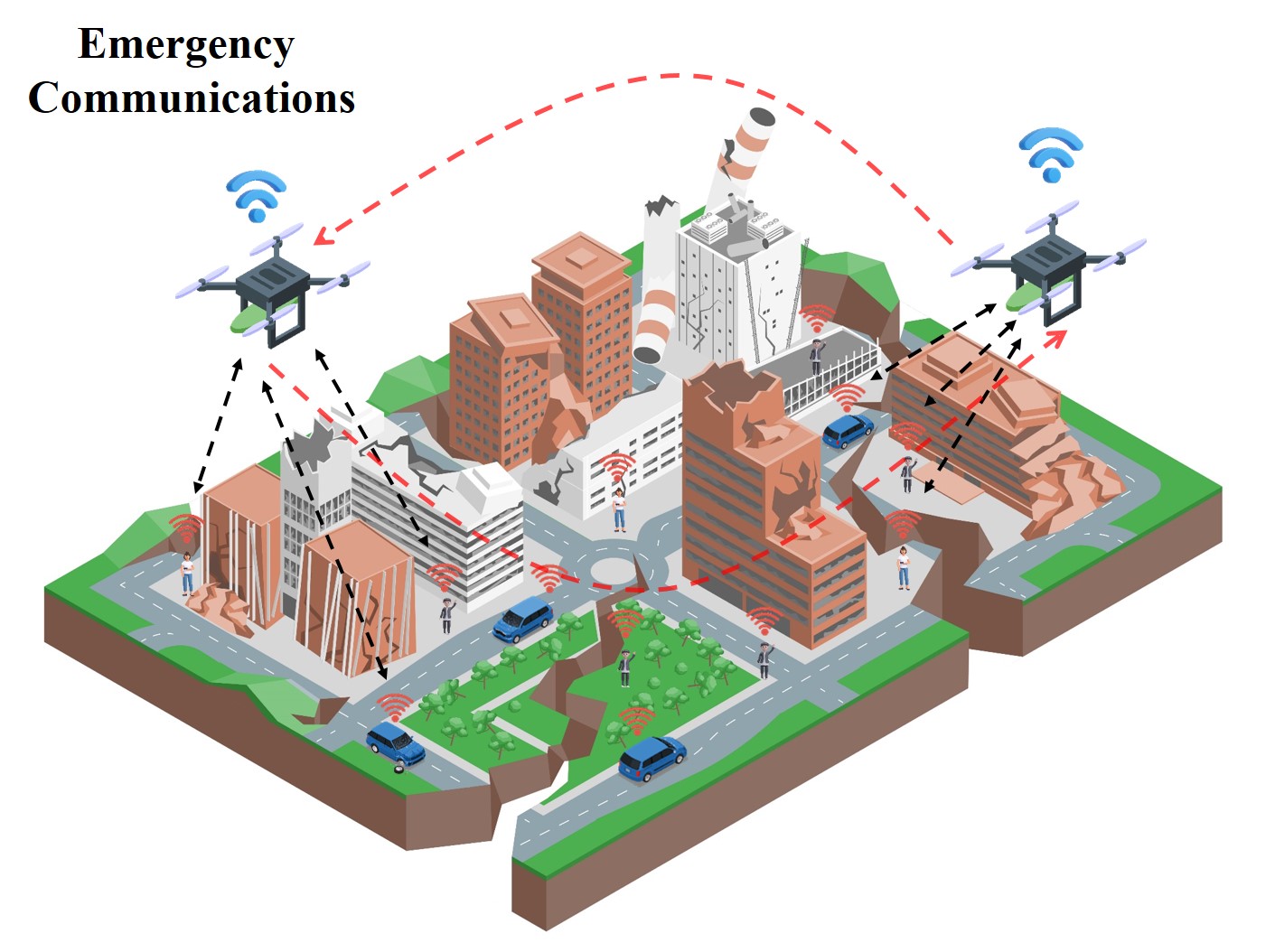}}
    
    \label{fig:2}
    }
     \subfigure[UAV networking]{
    \scalebox{0.24}[0.24]{\includegraphics{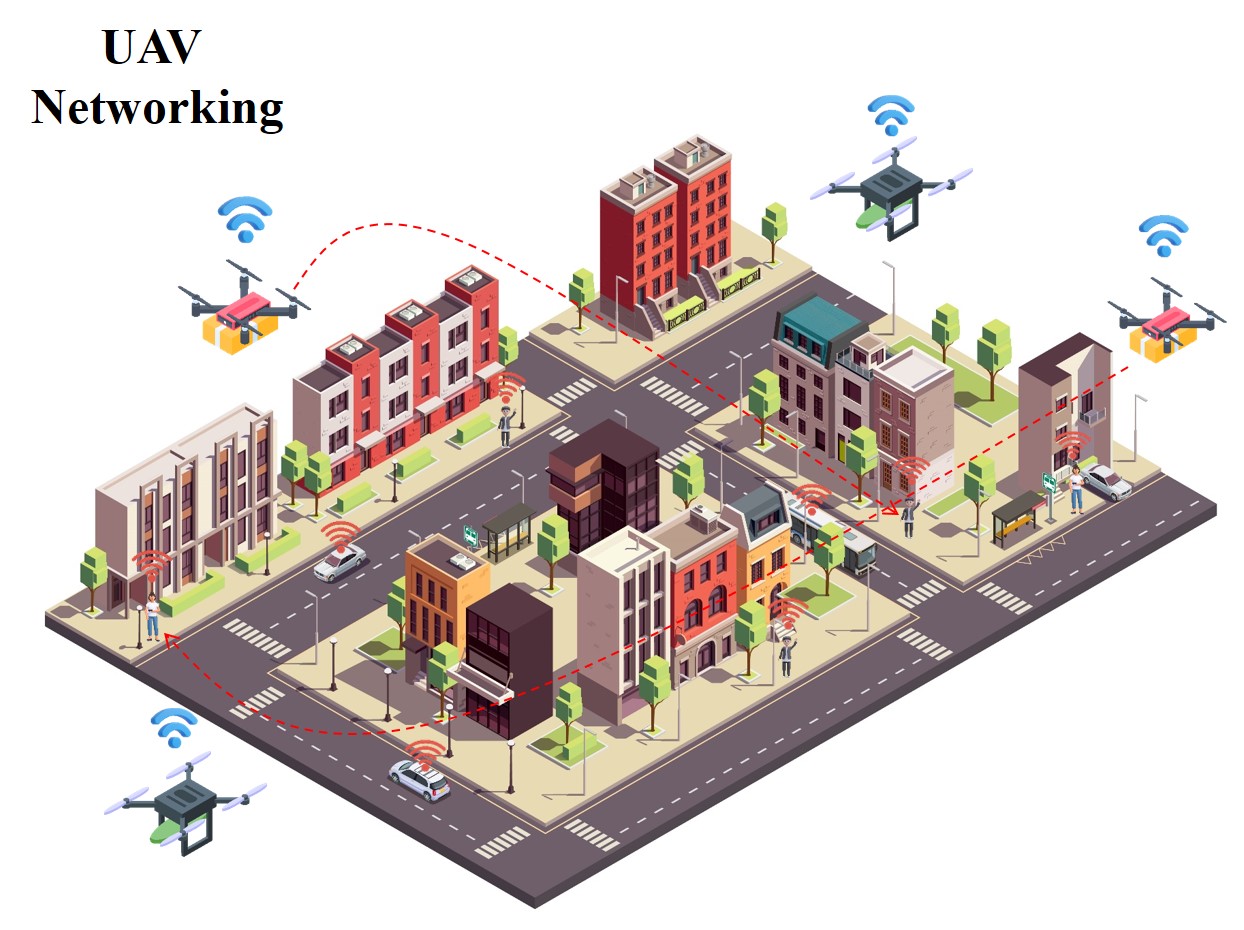}}
    \label{fig:3}
    }
    \subfigure[Environmental Monitoring]{
    \scalebox{0.23}[0.23]{\includegraphics{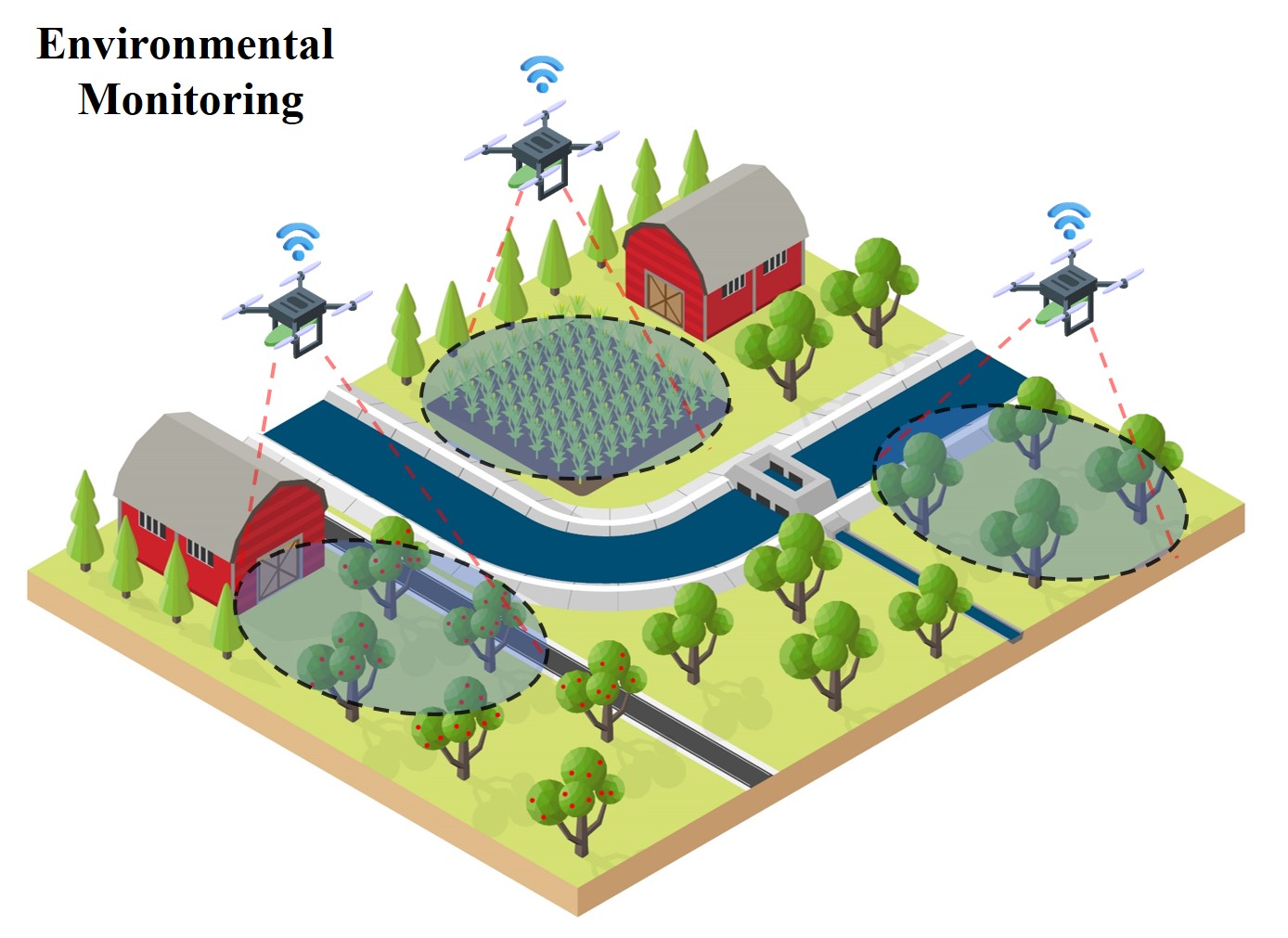}}
    
    \label{fig:2}
    }
    \caption{Three major applications of UAVs in the context of the low-altitude economy. In emergency communication scenarios, UAVs mainly serve as mobile base stations to provide communication services to users in emergency scenarios. In UAV networking, UAVs can communicate with vehicles and traffic management systems to improve the safety and efficiency of urban transportation, as well as communicate with terminal devices and control and command UAVs to perform tasks. In environmental monitoring, UAVs can collect real-time data to help governments and research institutions with environmental protection and resource management.}
    
  \end{center}
  \vspace{-0.6cm}
\end{figure*}

These resource allocation challenges in real-time and high-reliability wireless communication networks are mathematically formulated as mixed-integer non-convex problems. Traditional optimization techniques typically employ approximate convex optimization or iterative optimization. For instance, in addressing the channel resource allocation problem in UAV networks, a hybrid approach combining matching exchange theory and continuous convex optimization has been proposed to solve \cite{R4-2}. Similarly, an efficient iterative algorithm has been developed to optimize both communication resource allocation and UAV trajectory in dual-mobile environments involving fixed-wing UAVs and mobile users \cite{R4-3}. However, these conventional optimization methods rely on precise mathematical models and predefined problem structures, which constrain their adaptability to dynamically changing wireless communication environments. To overcome these limitations, researchers have increasingly explored artificial intelligence(AI)-based optimization strategies to enhance the adaptability and generalization of resource allocation solutions in wireless communication networks \cite{R5, R6, R7, R8}. Notably, deep reinforcement learning (DRL) methods have demonstrated significant advantages in addressing complex, high-dimensional problems. DRL can dynamically adjust policies in changing environments, ensuring robustness and effectively managing uncertainties and noise within the system. However, most DRL-based solutions for wireless resource allocation rely heavily on online deployment, which demands considerable training time. The practicality of the online DRL model is limited when the target deployment environment is unavailable or when stringent time constraints restrict initial training. Additionally, since DRL models are typically optimized based on the current state distribution of the wireless environment, they require additional adaptation time to adjust when environmental conditions change. Moreover, the existing online DRL optimization schemes struggle to simultaneously meet the requirements of unavailable environments and diverse target scenarios.

Optimized trajectory generation under the DRL optimization model is a potential solution to overcome the above problems. In recent years, Generative Artificial Intelligence (GAI) has proven particularly effective in modeling complex data distributions and generating high-quality samples \cite{R9,RY-1}. Generative algorithms such as Generative Adversarial Networks (GAN), Variational Autoencoders (VAE), and Diffusion Models (DM) have garnered widespread attention in the field of wireless communications\cite{R9-1,R9-2,R10,R11}. GANs were utilized to generate random sample models that mimic the distribution of real reflection channels, thereby assisting in efficient Intelligent Reflecting Surface (IRS) channel modeling\cite{R9-1}. In studies focusing on system robustness within semantic communication, the VAE method was employed to reconstruct image mask data, mitigating the impact of semantic noise on data transmission accuracy\cite{R9-2}. However, DMs exhibit superior algorithmic performance in terms of training stability, architectural flexibility, and theoretical foundation, which enhances their robustness in generative tasks. Consequently, researchers embark on integrating DMs with DRL to address the optimization problem in wireless resource allocation\cite{R10,R11}. Moreover, the design paradigm of Artificial Intelligence-Generated Networks (AIGN) enables the rapid generation of various customized network solutions, achieving expert-free problem optimization\cite{R12}. Therefore, leveraging the creative generation ability of Generative DM (GDM) can overcome the shortcomings of online DRL methods in practical applications. In an offline manner, the generated high-quality optimization trajectories can be used to build a more stable and highly generalized resource allocation optimization model.

In addition, conditional generation algorithms based on DM models have been widely used to guide the generation of target images \cite{RY-2,b2}, which can effectively solve the problem that the initial DM algorithm cannot complete directional generation. Therefore,  our core is to focus on constructing the feature structure of these conditions so that the conditional features can be incorporated into the resource allocation optimization problem. Knowledge graphs have been proven to effectively store and display real entities, relationships, and attributes in graphics \cite{R14}, and have strong application capabilities in knowledge reasoning and visual analysis \cite{b1}. Therefore, our work considers the design of the wireless network intent (WNI) paradigm to guide the generation of resource allocation optimization trajectories of the target network. \emph{The WNI in this paper can be defined as the objective factors existing in the wireless network that affect the system performance. For example, in the low-altitude economic application scenarios shown in Fig. 1, the intention can be abstractly defined as the channel conditions, application requirements, and optimization targets, etc. The specific design paradigm will be discussed in detail in section III-A.}

Given the above background and challenges, this paper proposes a WNI-guided trajectory generation optimization network based on cross-attention and GDM fusion for wireless communication systems. The detailed contributions of this paper are summarized as follows:

\begin{itemize}
\item We analyze the correlation between optimization trajectory data and GDM applications in wireless communication systems, constructing a background knowledge base (BKB) suitable for wireless resource optimization. This knowledge base stores the channel state distribution that aligns with the intention of the wireless communication structure, supporting GDM in generating the intent network.
\item	We propose an intent generation network based on cross-attention and GDM. The distribution of intent optimization trajectory tuples is matched with the respective intent labels based on cross-attention. Subsequently, we integrate a
multi-head attention mechanism and improve the underlying
Multi-layer Perceptron (MLP) network structure. The structure is fused to guide GDM in generating intent trajectories, which is named attention-MLP (AMLP). Additionally, different policy gradients and generation strategies for resource allocation optimization of trajectory tuples are designed and defined.
\item	With the aim of superior and stable system performance and low online training overhead, we combine the WNI-guided trajectory generation model with the offline DRL model to construct an optimization strategy for the WNI target network and solve the wireless resource allocation problem. This method leverages intent-guided creative data generation and offline learning capabilities to solve the aforementioned challenges.

\end{itemize}

The rest of the paper is organized as follows: Section II describes the details of the system model. Section III presents the proposed algorithm. Section IV reports and discusses the simulation results. Finally, Section V presents conclusions and future work.

\section{System Model and Problem Formulation}

\subsection{Wireless Network Intent-guided Trajectory Generation System Model}

\begin{figure*}[]
  \begin{center}
    \scalebox{0.75}[0.75]{\includegraphics{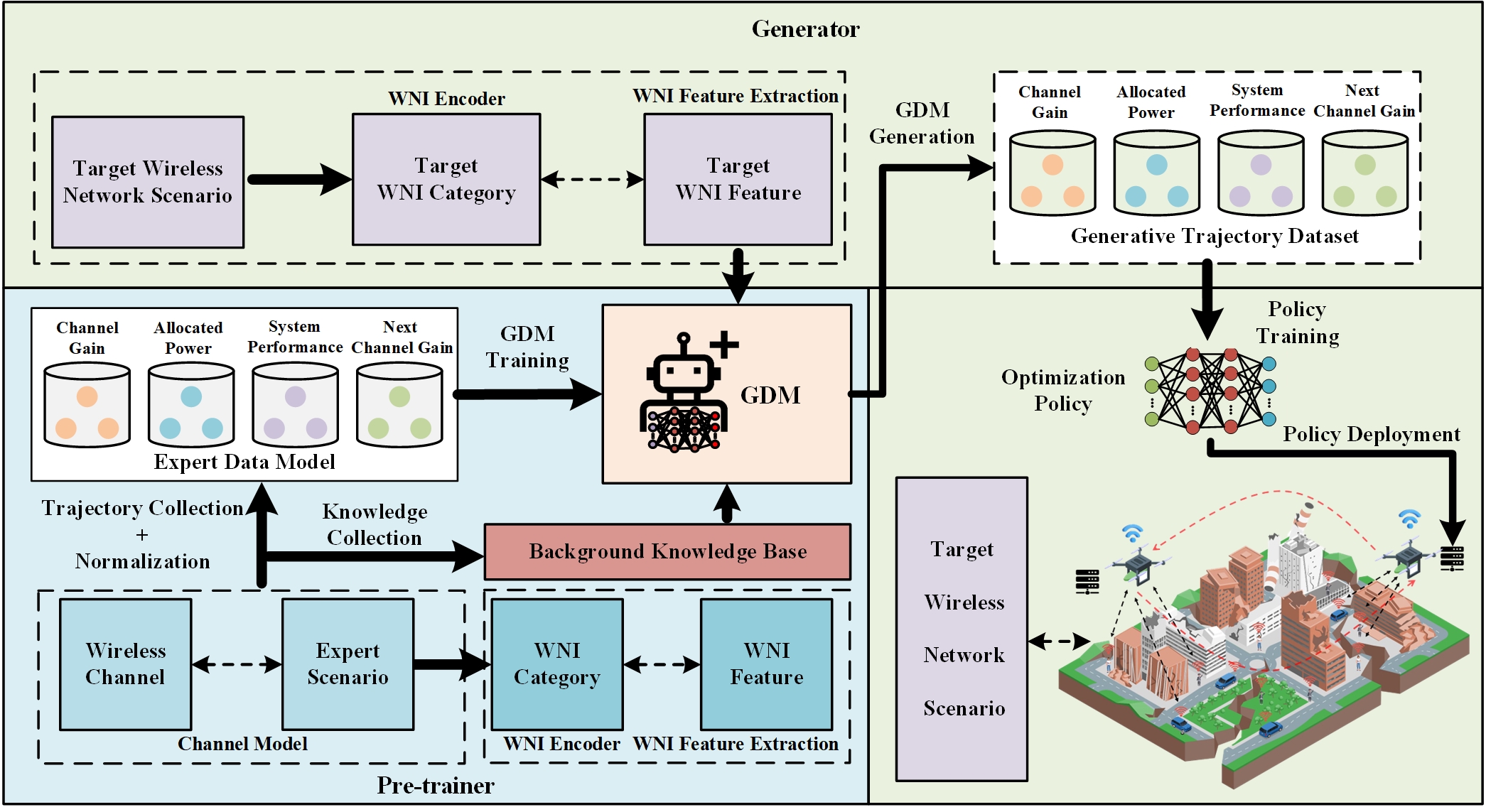}}
    \caption{Considered WNI-guided trajectory generation system model for wireless resource allocation optimization.}
    \label{fig:1}
  \end{center}
\end{figure*}

This paper considers a WNI-guided DRL optimization trajectory generation wireless communication system for resource allocation optimization. The demand side only needs to provide the entity attribute intent features of the wireless communication system, referred to as WNI in this paper, to the cloud generation model. This model can then generate resource allocation optimization trajectories that conform to the structure of the intentional wireless system. Importantly, this process does not require online deployment or model learning for wireless resource allocation optimization. This method will provide a design paradigm for solving the limitations of online AI resource allocation optimization in the low-altitude economy and other complex wireless network scenarios through offline creation and generation. Specifically, we refer to the wireless network generation paradigm of \cite{R12} for a more detailed design (as shown in Fig. 2), which mainly includes a pre-trainer and generator. The pre-trainer mainly guides the training process of WNI-guided trajectory GDM through an expert data model and BKB, while the generator generates a large amount of trajectory data through GDM and target WNI, and provides optimization policies to the target network. The core modules of this system are as follows:


\textbf{WNI Encoder:} The knowledge graph is used to extract entity attribute intention features from the wireless network structure. This intention feature covers the entity attribute structure intention in a cell, which will be discussed in Section III. The encoded WNI $\Delta$ of the cell $C$ can be expressed as:
\begin{equation}
    \Delta=E_{WNI}(C,\omega)
\end{equation}
where $E_{WNI}(\cdot)$ is the knowledge graph encoding network, $\omega$ is the network parameter.

\textbf{Channel Model:}  The factors influencing wireless network performance span multiple layers, encompassing user requirements, channel conditions, infrastructure design, and protocol stack parameters. Additionally, dynamic factors such as noise, interference, and user mobility continuously impact the quality of communication services. However, these variations can be abstractly represented by the channel matrix $H(\Delta)$, which reflects the channel state in different WNI $\Delta$. Consequently, the network model can be represented as:
\begin{equation}
    y_{\Delta_{i}} = H(\Delta_{i}) \cdot x_{\Delta_{i}} + n
\end{equation}
where $\Delta_{i}$ denotes $i$-th WNI, and $H(\Delta_{i})$ denotes the channel coefficient characterizes under WNI $\Delta_{i}$. The channel gain of the signal, represented as $g_{\Delta_{i}} = \mathbb{E}[|H(\Delta_{i})|^2]$. Without loss of generality, we define multiple independent distributions of the channel gain $g_{\Delta_{i}}$ to generalize the performance impact of the channel coefficient $H(\Delta_{i})$ under various WNIs $\Delta$.


\textbf{Expert Data Model:} For the original cell $\mathbf{C}=\{C_{1},C_{2},\cdots,C_{M}\}$ scenario with $M$ WNI feature sets $\mathbf{\Delta}=\{\Delta_1,\Delta_2,\cdots,\Delta_M\}$, the downlink expert trajectory tuple for each user in the scenario $C$ is used as the training dataset for the generative model. For each WNI $\Delta_{i}$ scenario, a total of $K$ trajectory data are collected, one of which is denoted as $\tau_{j}^{\Delta_{i}}=(s_{j}^{\Delta_{i}},a_{j}^{\Delta_{i}},r_{j}^{\Delta_{i}},s_{j}^{\prime\Delta_{i}})$. In the above context of the channel model, the trajectory elements $s_{j}^{\Delta_{i}}$, $a_{j}^{\Delta_{i}}$, $r_{j}^{\Delta_{i}}$, and $s_{j}^{\prime\Delta_{i}}$ correspond to the channel gain $g_{j}^{\Delta_{i}}$, the allocated power $p_{j}^{\Delta_{i}}$, the target system optimization performance $\mathcal{G}_{j}(\Delta_i)$, and the next time channel gain $s_{j}^{\prime\Delta_{i}}$, respectively. Then the expert dataset corresponding to the intention $\Delta_i$ can be expressed as $\Delta_{i}\leftrightarrow D_{\Delta_{i}}=\{\tau_{1}^{\Delta_{i}},\tau_{2}^{\Delta_{i}},\cdots,\tau_{j}^{\Delta_{i}},\cdots\tau_{K}^{\Delta_{i}}\}$, and the expert dataset of the entire intention $\Delta$ is denoted as $\mathbf{D}=\{D_{\Delta_{1}},D_{\Delta_{2}},\cdots,D_{\Delta_{M}}\}$. From a mathematical perspective, the training and generation of the DM model rely on the  $N{\sim}(0,1)$ Gaussian distribution, so the construction of the data model $\mathbf{D}$ must also meet this condition. To intuitively reflect the differences between different WNIs, we assume that the distribution of $D_{\Delta_{i}}$ corresponding to different WNI $\Delta_{i}$ in the entire $\mathbf{D}$ is independent of each other, that is, the joint probability density function satisfies
\begin{multline}
    f_{\mathbf{D}}(\tau^{\Delta_{1}},\tau^{\Delta_{2}},\cdots,\tau^{\Delta_{M}}) = \\
    f_{D_{\Delta_{1}}}(\tau^{\Delta_{1}}) \cdot f_{D_{\Delta_{2}}}(\tau^{\Delta_{2}}) \cdots f_{D_{\Delta_{M}}}(\tau^{\Delta_{M}}).
\end{multline}
The boundary of $\mathbf{D}$ in the Gaussian distribution is between $(\alpha,\beta)$. Then the distribution examples of different WNI data in the dataset can be shown in Fig. 3, where each area represents the trajectory distribution range of different WNI under the Gaussian distribution.

\begin{figure}
   \begin{center}
    \scalebox{0.48}[0.55]{\includegraphics{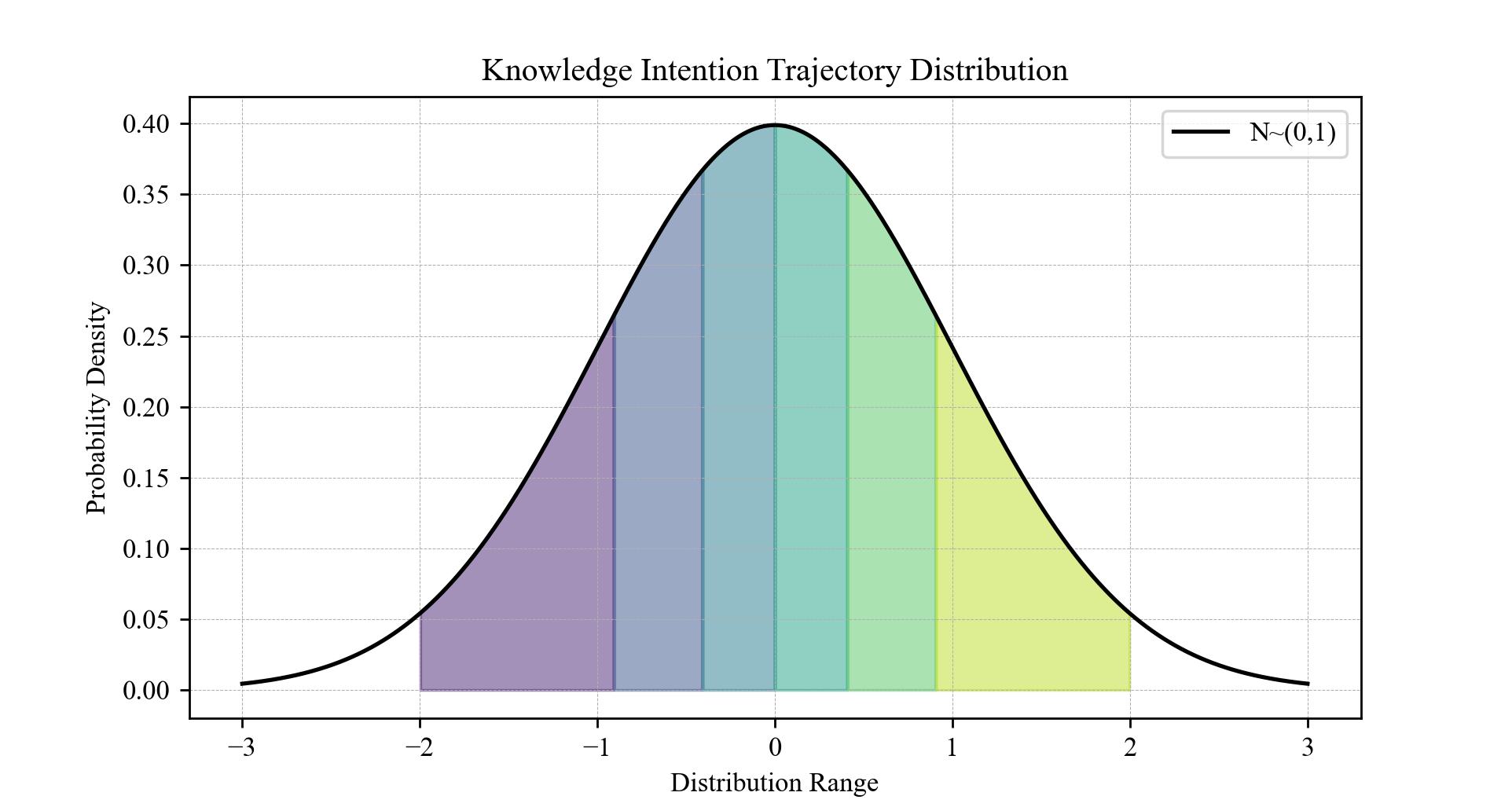}}
    \caption{An example for WNI-based data model distribution.}
    \label{fig:2}
  \end{center}
  \vspace{-0.6cm}
\end{figure}

In addition, data generation process may result in the generated trajectory data being outside the distribution range of expert data, which will greatly reduce the reliability of the generated data. Therefore, the guidance of the distribution range of the intent network trajectory during the generation process will improve the credibility of the generated results.
Based on the relationship between various network intentions and optimization trajectory distributions, the mathematical function $f(\cdot)$, which captures this relationship, is stored in the cloud-based BKB, denoted as ${\mathcal{B}}$ \cite{R13}. This function primarily includes the mapping required for optimizing trajectory normalization and the numerical distribution ranges corresponding to different WNI. The stored knowledge serves as guidance for the training and generation processes of the generative model.

\textbf{WNI-guided Trajectory Generation Diffusion Model:} The WNI-based expert data \textbf{D} are used to support offline generative model training of $G(\cdot)$, and the trained generative model $G(\cdot)$ of the system is deployed in the cloud. When the generative model obtains the WNI $\Delta$, $G(\cdot)$ will generate the corresponding optimized trajectory data as follows:
\begin{equation}
\tilde{s}_j^{\Delta_i}=G(\Delta_i,N\sim(0,1);{\theta}_s|{\mathcal{B}})
\end{equation}

\begin{equation}
    \tilde{a}_{j}^{\Delta_{i}}= G(\Delta_{i},N{\sim}(0,1),\tilde{s}_{j}^{\Delta_{i}};\theta_{a}|{\mathcal{B}})
\end{equation}

\begin{equation}
    \tilde{r}_{j}^{\Delta_{i}}= G(\Delta_{i},N{\sim}(0,1),\tilde{s}_{j}^{\Delta_{i}},\tilde{a}_{j}^{\Delta_{i}};\theta_{r}|{\mathcal{B}})
\end{equation}

\begin{equation}
    \tilde{s}_{j}^{\prime\Delta_{i}}= G(\Delta_{i},N{\sim}(0,1),\tilde{s}_{j}^{\Delta_{i}},\tilde{a}_{j}^{\Delta_{i}},\tilde{r}_{j}^{\Delta_{i}};\theta_{s^{\prime}}|{\mathcal{B}})
\end{equation}
where ${\theta}_s$, ${\theta}_a$, ${\theta}_r$ and ${\theta}_s^{\prime}$ correspond to the trainable parameters of the state, action, reward and next state generation model $G(\cdot)$, respectively. $N$ is random Gaussian noise, and $\mathcal{B}$ indicates that BKB is used to assist data generation. Then a generated optimization trajectory can be recorded as $\tilde{\tau}_{j}^{\Delta_{i}}=(\tilde{s}_{j}^{\Delta_{i}},\tilde{a}_{j}^{\Delta_{i}},\tilde{r}_{j}^{\Delta_{i}},\tilde{s'}_{j}^{\Delta_{i}})$. $\tilde{\tau}^{\Delta_{i}}$ will be used for offline optimization model training of wireless network under intention $\Delta_{i}$ to assist in solving the resource allocation optimization problem of wireless network.

\subsection{Problem Formulation}
The method proposed in this paper aims to address the general problem of resource optimization in wireless communication. To facilitate discussion and validation, we employ the classical power allocation model as an illustrative example. This model serves solely to demonstrate the effectiveness of the proposed method and does not restrict its potential applications. Accordingly, without loss of generality, we consider a base station with a total transmission power $P$, serving a group of users through multiple orthogonal channels \cite{R9}. In the communication model constructed in this paper, we assume that WNI $\Delta$ only controls the channel gain distribution of networks with different intentions, so that the channel gain data under different intentions can conform to the distribution in Fig.2. In this model, $g_{m}^{\Delta_{i}}$ represents the channel gain of the $m$-th channel in the $i$-th WNI $\Delta_{i}$ network environment, $p_{m}^{\Delta_{i}}$ represents the power allocated to the channel, and $n_{0}$ represents the power of additive white Gaussian noise. The optimization goal of the expert network is to maximize the total spectral efficiency $\mathcal{G}(\Delta_i)$ by optimizing the power allocation between channels under different WNI $\Delta_i$, as follows:









\begin{subequations}
\begin{align}
\max_{\{p_1^{\Delta_i},...,p_M^{\Delta_i}\}}\quad &\mathcal{G}(\Delta_i)=\sum_{m=1}^M log_2 (1+\frac{g_m^{\Delta_i}p_m^{\Delta_i}}{n_0}) \tag{8}\\[10pt] 
{s.t.} \quad & p_{m}^{\Delta_{i}} \geq 0, \forall m  \\[10pt]
&\sum_{m=1}^Mp_m^{\Delta_i}\leq P^{\Delta_i}\\[10pt] &\Delta_{i} \in \mathbf{\Delta}\\[10pt] &\alpha_s^{\Delta_i}\leq f_{D_{\Delta_i}}(g^{\Delta_i})\leq\beta_s^{\Delta_i}\\[10pt] &\alpha_{a}^{\Delta_{i}}\leq f_{D_{\Delta_{i}}}(p^{\Delta_{i}})\leq\beta_{a}^{\Delta_{i}}
\end{align}
\end{subequations}
where $P^{\Delta_{i}}$ is the total power of the  intent $\Delta_{i}$, $f_{D_{\Delta_i}}(g^{\Delta_i})$, $ f_{D_{\Delta_{i}}}(p^{\Delta_{i}})$, represent the data distribution range of the wireless network intent $\Delta_{i}$ channel gain $g^{\Delta_i}$ and power $p^{\Delta_{i}}$ respectively in $[\alpha_{s}^{\Delta_{i}},\beta_{s}^{\Delta_{i}}]$ and $[\alpha_{a}^{\Delta_{i}},\beta_{a}^{\Delta_{i}}]$, and $\mathbf{\Delta}$ is the set of all network intents. Constraint (8a) defines the minimum limit of the power allocated to each user, constraint (8b) ensures that the allocated power does not exceed the total power, Constraint (8c) ensures that the model is applied within the WNI set $\mathbf{\Delta}$, and constraints (8d) and (8e) ensure that the data are within the Gaussian distribution range of the corresponding intent, thereby ensuring the accuracy between the data and the intent.

In the aim, the optimization goal of the system network is to optimize the power allocation between channels under the different WNI $\Delta$ through the trajectory generation capability of GDM $G(\cdot)$ and maximize the target communication system performance $\mathbf{y}_{\Delta_{i}}$. Then the optimization target of this paper can be abstracted from (8) and adjusted as follows:

\begin{equation}
    \mathcal{J}(\Delta)=E_{OFF}({G}(\Delta;\theta_s,\theta_a,\theta_r,\theta_{s^{\prime}}|\mathcal{B});\varpi)
\end{equation}

\begin{equation}
    \max_{\{\tilde{p}_{1}^{\Delta_{i}},...,\tilde{p}_{M}^{\Delta_{i}}\}}{y}_{\Delta_{i}}=\sum_{m=1}^{M}H(\mathcal{J}(\Delta_{i})){x}_{\Delta_{i}}+{n}
\end{equation}
where $E_{OFF}(\cdot)$ is an offline encoding network based on the generated intent trajectory, $\varpi$ is a trainable parameter of the optimization encoding network, and $\mathcal{J}(\cdot)$ represents the intent network optimization model based on the wireless network intent generation trajectory. $H(\cdot)$ is the intent network channel matrix, which relies on $\mathcal{J}(\cdot)$ to estimate the intent network channel state, and ultimately makes the entire intent network performance ${y}_{\Delta_{i}}$ optimal.


 \section{The Pipeline of WNI-guided Trajectory Generation System Construction}

In this section, we provide a detailed description of the WNI design for wireless communication networks. Following this, we explain the construction process of the WNI-guided and cross-attention DM based generative model. Finally, we elaborate on the development of the WNI wireless communication network optimization model based on generated optimization trajectories.

\subsection{Wireless Network Intention Design Paradigm}

\begin{figure}
   \begin{center}
    \scalebox{0.45}[0.45]{\includegraphics{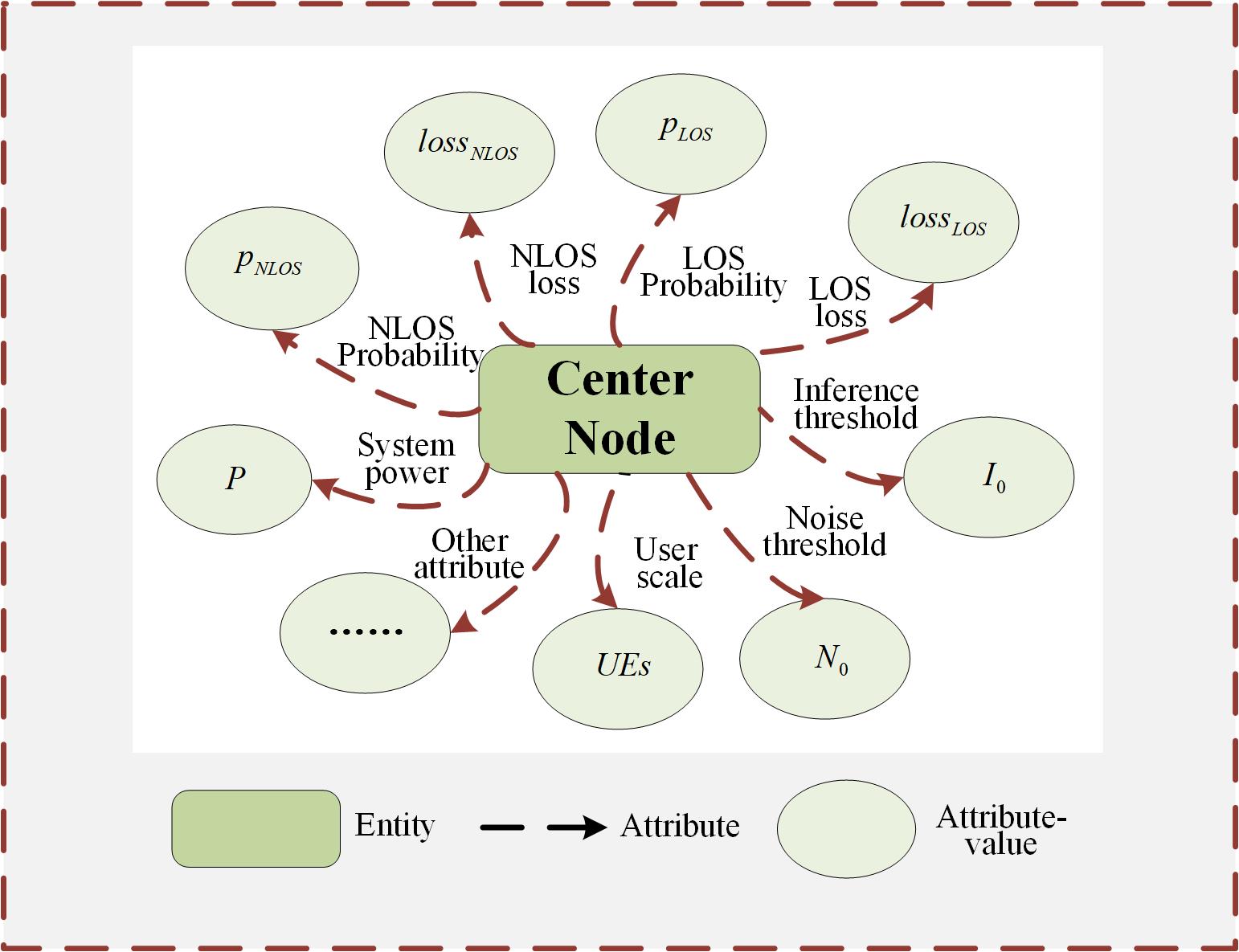}}
    \caption{WNI construction of wireless communication scenario.}
    \label{fig:2}
  \end{center}
  \vspace{-0.6cm}
\end{figure}

The WNI can be represented by the entities ($e$), attributes ($a$), and attribute values ($v$) of the directed graph related to the target communication scenario, which is viewed from the perspective of resource allocation optimization architecture \cite{R14}. The proposed WNI is based on the structural framework of the knowledge graph, which begins from the actual scenario and optimization targets, and aims to serve as a generation guide for the intent-guided trajectory generation model. WNI captures the deep semantic information of the target scenario from the communication architecture, and WNI captures the deep semantic information of the target scene from the communication architecture, where the communication structure is mainly composed of wireless network structure, channel conditions, and user characteristics. Its fundamental form considers the construction mode of the knowledge graph, guided by the target problem $\mathcal{G}(\cdot)$, as illustrated in Fig. 4.

The knowledge graph feature structure of WNI $\Delta$ can be represented as an entity-attribute-value tuple associated with $\mathcal{G}(\cdot)$. The WNI in Fig. 4 can be represented by the feature $\mathbf{E}_{\Delta}$ of directed graphs as follows:

\begin{figure}[]
  \begin{center}
  
  \subfigure[Multi-head attention mechanism]{
    \scalebox{0.6}[0.6]{\includegraphics{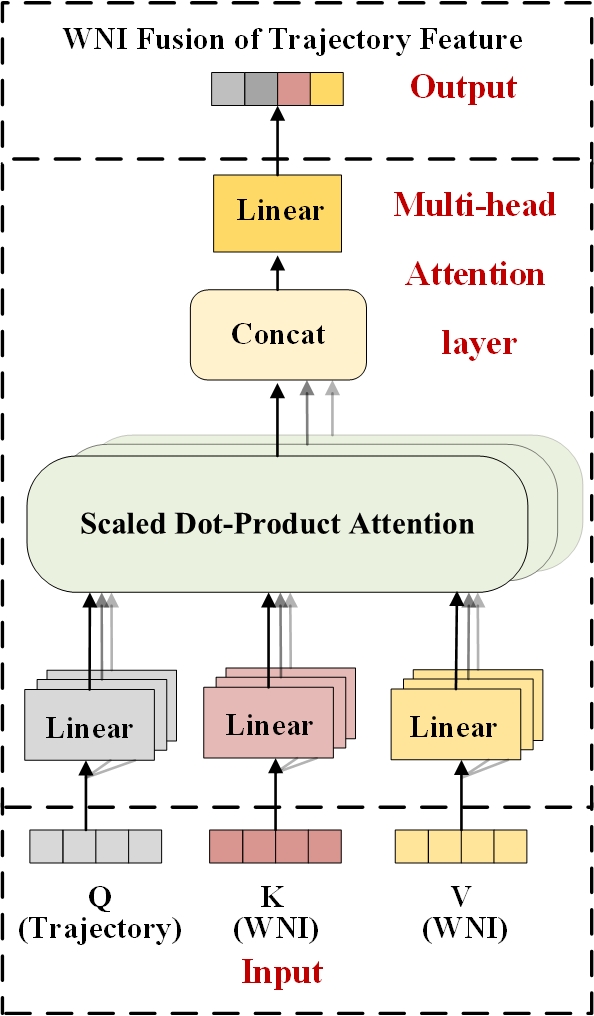}}
    
    \label{fig:2}
    }
     \subfigure[Attention-MLP Network]{
    \scalebox{0.6}[0.56]{\includegraphics{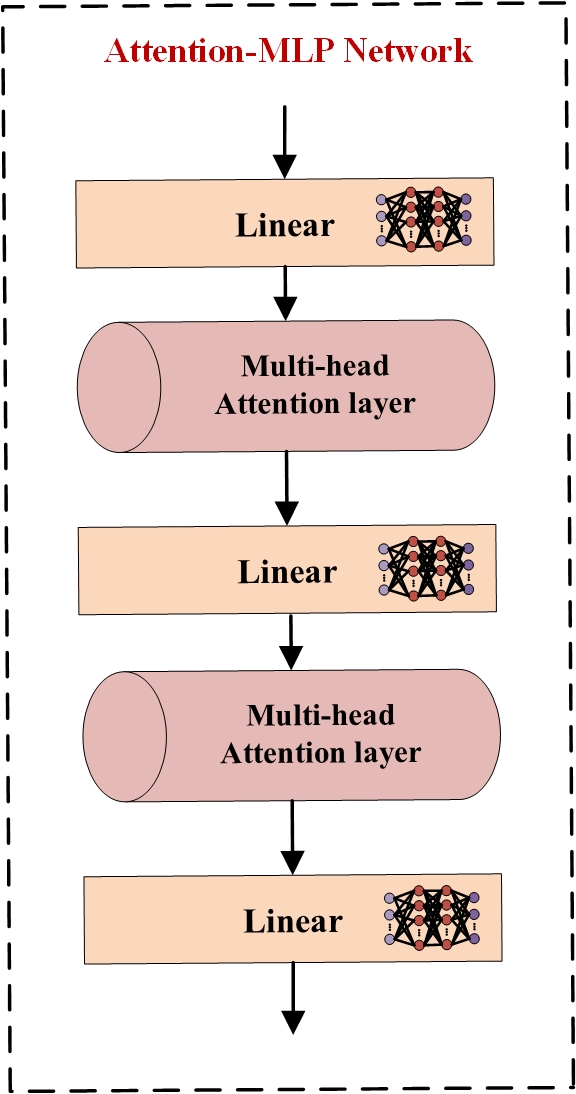}}
    \label{fig:3}
    }    \caption{The structure of the attention-MLP network for WNI-guided trajectory generation.}
  \end{center}
  \vspace{-0.6cm}
\end{figure}

\begin{figure*}[]
  \begin{center}
  
  \subfigure[WNI-guided Trajectory  Generation model adding noise for training ]{
    \scalebox{0.23}[0.22]{\includegraphics{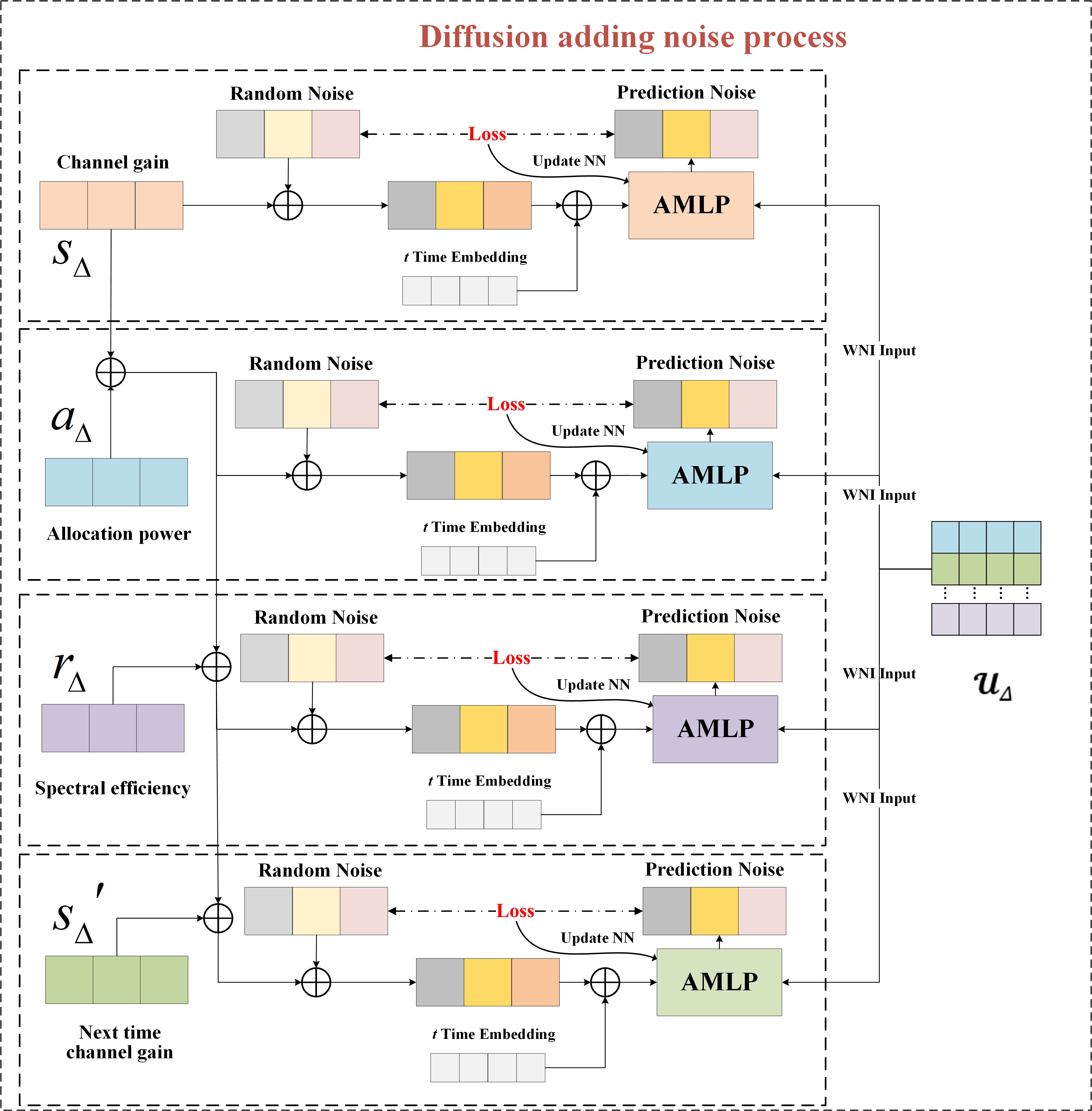}}
    
    \label{fig:2}
    }
     \subfigure[WNI-guided Trajectory Generation model denoising for generation]{
    \scalebox{0.23}[0.186]{\includegraphics{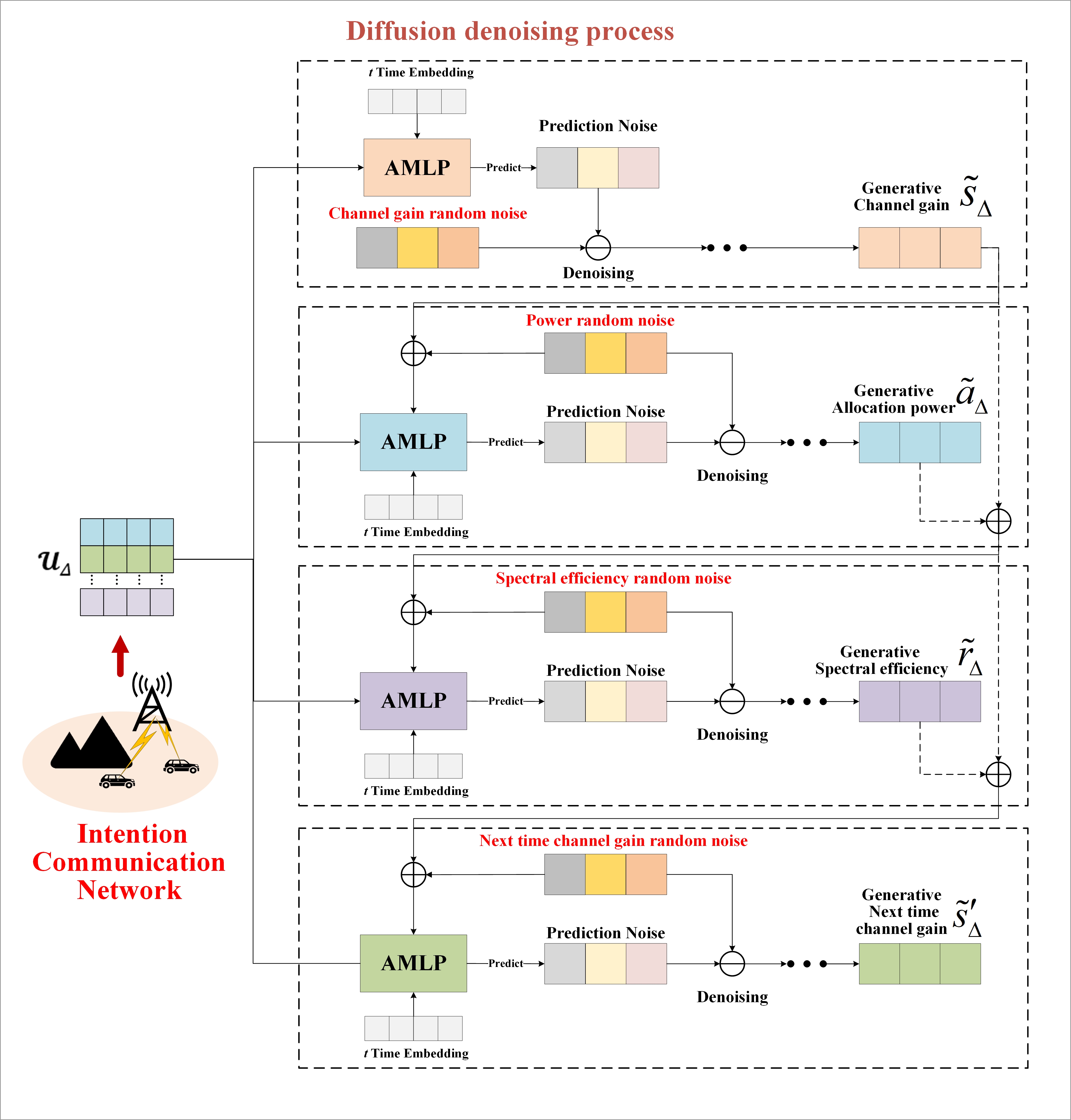}}
    \label{fig:3}
    }    \caption{The framework of the WNI-guided trajectory generation model.}
  \end{center}
  \vspace{-0.6cm}
\end{figure*}

\begin{equation}
    \boldsymbol{E}_{\Delta}=\left\{\left.\left(\begin{array}{ccc}
e_{\Delta} & a_{\Delta}^{1} & v_{\Delta}^{1} \\
e_{\Delta} & a_{\Delta}^{2} & v_{\Delta}^{2} \\
\vdots & \vdots & \vdots \\
e_{\Delta} & a_{\Delta}^{N} & v_{\Delta}^{N}
\end{array}\right) \right\rvert\, e \in \mathcal{E}, a \in \mathcal{L}\right\}
\end{equation}
where $e_{\Delta}$ denotes the central node of the target communication scenario, $a_{\Delta}^{i} $ denotes the $i$-th attribute relationship, and $v_{\Delta}^{i}$ denotes the attribute range threshold under the $i$-th attribute relationship. $\mathcal{E}$ denotes the central node set,  such as $\mathcal{E}=\{\text{UAV},\text{BS},\text{AP},\ldots\}$, which act as control centers for network resources and are responsible for the dynamic allocation of resources, including spectrum and power. $\mathcal{L}$ denotes the relationship attribute set. The construction of relationship attributes can be defined as $\mathcal{L}=\{$ User scale, Noise, Inference, LOS loss, LOS probability, NLOS loss, NLOS probability, Transmission power, $\ldots\}$ from the optimization architecture.

In addition, we only focus on the attribute $a$ and attribute value $v$ of the directed graph feature $\boldsymbol{E}_{\Delta}$ for the target communication scenario of a single node $e_{\Delta}$, which can enable a deep semantic feature representation of WNI $\Delta$. Therefore, $E_{\Delta}$ can be transformed into a WNI feature vector $\mathcal{U}_{\Delta}$ by embedding the attribute $a$ and the attribute value $v$ using word embedding, as follows:
\begin{equation}
\mathcal{U}_{\Delta}=\begin{bmatrix}\boldsymbol{h}_{a_{\Delta}^{1}}&\boldsymbol{h}_{\nu_{\Delta}^{1}}\\\boldsymbol{h}_{a_{\Delta}^{2}}&\boldsymbol{h}_{\nu_{\Delta}^{2}}\\\vdots&\vdots\\\boldsymbol{h}_{a_{\Delta}^{N}}&\boldsymbol{h}_{\nu_{\Delta}^{N}}\end{bmatrix}
\end{equation}
where $\boldsymbol{h}_{a_{\Delta}^{i}}$ and $\boldsymbol{h}_{v_{\Delta}^{i}}$ represent the word vector features of the $a_{\Delta}^{i}$ attribute and the $v_{\Delta}^{i}$ attribute value, respectively. The intent feature $\mathcal{U}_{\Delta}$ after WNI-Encoder encodes will be used to establish an intent association with the expert trajectory collected from the target communication scenario and also used for the training and generation of the WNI-guided trajectory generation model.

\subsection{Wireless Network Intention-guided Trajectory Generation Model for Resource Allocation Optimization}

In this paper, the DRL agent and the communication system engage in a Markov decision process (MDP) to optimize resource allocation within a regular interaction interval $t$. The channel state $s_{\Delta}$, optimization action $a_{\Delta}$, system performance $r_{\Delta}$, and the next channel state $s'_{\Delta}$ during system interaction for the $\mathcal{G}(\Delta)$ problem are denoted as the optimization trajectory. The expert trajectory $\mathbf{D}$ accurately characterizes the system dynamic relation, enabling the maximization of long-term expected rewards across different target scenarios $\Delta$ within a complete wireless task cycle $T$. Furthermore, in relation to the target optimization problem $\mathcal{G}(\Delta)$, the trajectory in this paper is defined as follows.

\textbf{Action Space}: The definition of the action space is closely linked to the determination of actions for the optimization problem $\mathcal{G}(\cdot)$. The optimization strategy can be defined based on the specific optimization problem, which encompasses parameters such as aggregate frame length \cite{R5}, transmission power \cite{b3,b4}, and computation frequency \cite{R7}. Again, this paper focuses on the power allocation optimization problem within the WNI $\Delta$. Consequently, the action is represented as $a_{\Delta} = \left[p_{\Delta}^{1}, p_{\Delta}^{2}, \dots, p_{\Delta}^{N}\right]$, where $p_{\Delta}$ denotes the allocated power in the intention network $\Delta$.

\textbf{Observation Space}: SINR and RSSI are commonly employed as objective metrics to characterize the quality of communication services and are therefore widely used to define the states in DRL-based communication optimization problems \cite{R5, R7}. Channel gain plays a critical role in determining the transmission quality of the signal, and under fixed noise and interference levels, exhibits a positive correlation with SINR or RSSI. To ensure generalization, the paper considers that the variations in the WNIs $\Delta$ influence only the distribution of the channel gain $g_{\Delta}$. As a result, the DRL state can be represented as $s_{\Delta} = \left[g_{\Delta}^{1}, g_{\Delta}^{2}, \dots, g_{\Delta}^{N}\right]$.

\textbf{Reward}: Rewards are utilized to evaluate the quality of communication during interactions between agents and communication networks in a DRL-based communication system. The optimization objective of this paper is to maximize long-term spectrum efficiency $max \sum_{t=0}^{T} \mathcal{G}(\Delta)=\sum_{t=0}^{T} \gamma^{t} r_{\Delta }^{t}$. Therefore, the reward is defined as $r_{\Delta} = \left[\mathcal{G}(\Delta)^{1}, \mathcal{G}(\Delta)^{2}, \dots, \mathcal{G}(\Delta)^{N}\right]$ .

The DM \cite{R15} learns and generates the data distribution $p(x)$ by sequentially adding noise in a forward process and denoising in a reverse process along a fixed Markov chain over $T$ learning steps. A cross-attention mechanism is employed to enhance the underlying network encoder $\epsilon_\theta$ of the DM, enabling the condition to influence the noise generation process \cite{R16}. Consequently, We integrate a multi-head attention mechanism and modify the underlying MLP network structure, which is named AMLP and can equip the model with the ability to recognize network intent $\Delta$, as illustrated in Fig. 5. Specifically, for the resource allocation optimization trajectory $\tau$ and the communication network WNI $\Delta$, the features of $\tau$ and the WNI features $\mathcal{U}_{\Delta}$ are fused with weighted coefficients and mapped to the intermediate layer of the network via an attention mechanism, as follows:

\begin{subequations}
    \begin{equation}
    {Attention}(Q_\tau,K_\Delta,V_\Delta)={softmax}\left(\frac{Q_\tau K_\Delta^{\mathrm{T}}}{\sqrt{d}}\right)\cdot V_\Delta 
\end{equation}
\begin{equation} Q_{\tau}=W_{Q_{\theta}}^{(i)}\cdot\varphi_{i}(\tau)
\end{equation}
\begin{equation} K_{\Delta}=W_{K_{\theta}}^{(i)}\cdot flatten(\mathcal{U}_{\Delta})
\end{equation}
\begin{equation}
V_{\Delta}=W_{V_{\theta}}^{(i)}\cdot flatten(\mathcal{U}_{\Delta})
\end{equation}
\end{subequations}
where $\varphi(\cdot)$ is the AMLP intermediate layer representation that implements $\epsilon_\theta$, $flatten(\cdot)$ is a dimensionality reduction function used to match the dimension with the AMLP network layer, and $W_{Q_{\theta}}^{(i)}$, $W_{K_{\theta}}^{(i)}$, and $W_{V_{\theta}}^{(i)}$ denote the learnable projection matrices. The feature fusion of communication network optimization trajectory and WNI $\Delta$ in latent space will be achieved through the attention mechanism \cite{R16}, i.e. $\tau=(s,a,r,s^{\prime})\stackrel{Attention(\cdot)}{\Rightarrow}(s,a,r,s^{\prime}|\Delta)$.

During the model training process, $\tau_0 \sim q(\tau) | \tau \in D$ is sampled from the communication network optimization trajectory set $D $ and random Gaussian noise $\epsilon \sim \mathcal{N}(0, I)$ is introduced. Furthermore, the AMLP model $\epsilon_\theta$ is then employed to predict this noise. Each forward diffusion step involves calculating the gradient between the AMLP model's predicted noise and the sampled noise, followed by a gradient update. The training is concluded after $T$ diffusion steps \cite{R15}. In addition, considering that the Markov decision process makes the trajectory elements have a time series relationship, i.e.$s\stackrel{\pi_{\vartheta}}{\Rightarrow}a$,$(s,a)\stackrel{H}{\Rightarrow}r$,$(s,a,r)\stackrel{\mathbf{p}(s\prime|s,a)}{\Rightarrow}s^{\prime}$, where $\pi_{\vartheta}$ is the optimization policy of DRL model, $H$ is the channel matrix of wireless network, and $\mathbf{p}$ is the channel state transition probability. Therefore, the feature of the time series relationship between the elements of the communication optimization trajectory is considered in the generative learning process. The training model architecture of the intention-guided trajectory generation in this paper is shown in Fig. 6 (a). The generated training gradient for the trajectory tuple $\tau=(s,a,r,s'|\Delta)$ is as follows:
\begin{equation}
\nabla_{\theta_s}\xi_s=\nabla_{\theta_s}\|\epsilon-\epsilon_{\theta_s}\big(\sqrt{\bar{\alpha}_t}s_0+\sqrt{1-\bar{\alpha}_t}\boldsymbol{\epsilon},t,\Delta\big)\|^2
\end{equation}
\begin{equation}  \nabla_{\theta_{a}}\xi_{a}=\nabla_{\theta_{a}}\|\epsilon-\epsilon_{\theta_{a}}\Big(\sqrt{\bar{\alpha}_{t}}a_{0}+\sqrt{1-\bar{\alpha}_{t}}\epsilon,t,\Delta,s_{0}\Big)\Big\|^{2}
\end{equation}
\begin{equation} \nabla_{\theta_{r}}\xi_{r}=\nabla_{\theta_{r}}\|\epsilon-\epsilon_{\theta_{r}}\big(\sqrt{\bar{\alpha}_{t}}r_{0}+\sqrt{1-\bar{\alpha}_{t}}\epsilon,t,\Delta,s_{0},a_{0}\big)\|^{2}
\end{equation}
\begin{equation}
    \nabla_{\theta_{s}\prime}\xi_{s\prime}=\\ \nabla_{\theta_{s}\prime}\|{\epsilon}-{\epsilon}_{\theta_{s}\prime}\big(\sqrt{\bar{\alpha}_{t}}s^{\prime}_{0}+\sqrt{1-\bar{\alpha}_{t}}{\epsilon},t,\Delta,s_{0},a_{0},r_{0}\big)\|^{2}
\end{equation}
where $\nabla_{\theta_s}\xi_s$, $\nabla_{\theta_{a}}\xi_{a}$, $\nabla_{\theta_{r}}\xi_{r}$ and $\nabla_{\theta_{s}\prime}\xi_{s\prime}$ are the noise prediction policy gradients for trajectory elements $s$, $a$, $r$, $s\prime$ respectively. In addition, the detailed training process of the AMLP model for the WNI-guided trajectory generation model $G(\cdot)$ is shown in Algorithm 1.

\begin{algorithm}[!h]
\DontPrintSemicolon
  Initialize AMLP model $\epsilon_{\theta_s}$,$\epsilon_{\theta_a}$,$\epsilon_{\theta_r}$,$\epsilon_{\theta_s^{\prime}}$ of trajectory tuple\;
  \textbf{Input:} Expert trajectory of power allocation optimization $\mathbf{D}$, Expert WNI set $\mathbf{\Delta}$, Noise coefficient $\alpha_t$ varying with time $t$, mini-batch $F_1$, time steps $T$.\;
     \For{$m=1$ to $M_{1}$}    
        { 
            Sample mini-batch $\left(s_{f}^{\Delta_{i}},a_{f}^{\Delta_{i}},r_{f}^{\Delta_{i}},s_{f}^{'\Delta_{i}}\right)$ of $F_1$ optimization trajectory from $\mathbf{D}$:
            $\left\{\left(g_{f}^{\Delta_{i}},p_{f}^{\Delta_{i}},\mathcal{G}_{f}(\Delta_{i}),g_{f}^{'\Delta_{i}}\right)|\Delta_{i}\in\boldsymbol{\Delta},f=1,\cdots,F_1\right\}$\;
            Extract the WNI intent feature $\mathcal{U}(\Delta_{i})$ corresponding to the sampling trajectory according to Eq. (12)\;
            Random noise sampling $\epsilon{\sim}\mathcal{N}(0,{I})$\;
            Random time sampling $t{\sim}Uniform(\{1,\cdots,T\})$\;
             Calculate gradient $\nabla_{\theta_s}$,$\nabla_{\theta_a}$,$\nabla_{\theta_r}$,$\nabla_{\theta_s^{\prime}}$ according to Eqs. (14)-(17)\;
             Update AMLP network parameters $\theta_s$,$\theta_a$,$\theta_r$,$\theta_s^{\prime}$\;
        }       
 \textbf{Output:} Trained AMLP model $\epsilon_{\theta_s}$,$\epsilon_{\theta_a}$,$\epsilon_{\theta_r}$,$\epsilon_{\theta_s^{\prime}}$ for WNI-guided generative model $G(\cdot)$\;
\caption{WNI-guided trajectory generative model training for power allocation optimization}
\end{algorithm} 

\begin{figure*}[t]
\begin{equation}
    \tilde{s}_{t-1} = clip\left(\frac{1}{\sqrt{\alpha_{t}}}\left(s_{t}-\frac{1-\alpha_{t}}{\sqrt{1-\overline{\alpha}_{t}}}{\epsilon}_{\theta_{s}}(s_{t},t,\Delta^{\prime})\right)+\sigma_{t}\boldsymbol{z},\alpha_{s}^{\Delta^{\prime}},\beta_{s}^{\Delta^{\prime}}\right)
\end{equation}

\begin{equation}
    \tilde{a}_{t-1}=clip\left(\frac1{\sqrt{\alpha_t}}\left(a_t-\frac{1-\alpha_t}{\sqrt{1-\bar{\alpha}_t}}\epsilon_{\theta_a}(\alpha_t,t,\Delta^{\prime},s_t)\right)+\sigma_t\boldsymbol{z},\alpha_a^{\Delta^{\prime}},\beta_a^{\Delta^{\prime}}\right)
\end{equation}

\begin{equation}
    \tilde{r}_{t-1}=clip\left(\frac{1}{\sqrt{\alpha_{t}}}\left(r_{t}-\frac{1-\alpha_{t}}{\sqrt{1-\bar{\alpha}_{t}}}\epsilon_{\theta_{r}}(r_{t},t,\Delta^{\prime},s_{t},a_{t})\right)+\sigma_{t}\boldsymbol{z},\alpha_{r}^{\Delta^{\prime}},\beta_{r}^{\Delta^{\prime}}\right)
\end{equation}

\begin{equation}
    \tilde{s}'_{t-1}=clip\left(\frac{1}{\sqrt{\alpha_{t}}}\Bigg(s'_{t}-\frac{1-\alpha_{t}}{\sqrt{1-\bar{\alpha}_{t}}}\epsilon_{\theta_{s'}}\big(s'_{t},t,\Delta',s_{t},a_{t},r_{t}\big)\Bigg)+\sigma_{t}\boldsymbol{z},\alpha_{s'}^{\Delta'},\beta_{s'}^{\Delta'}\right)
\end{equation}
\vspace{-0.6cm}
\end{figure*}
\setcounter{equation}{23} 
\begin{figure*}[!ht]
    \begin{equation}
    y=\tilde{r}_{j}^{\Delta_{i}}+\gamma max\left[\lambda min_{\tilde{a}_{j}^{\Delta_{i}}}Q_{\theta_{j}^{\prime}}(\tilde{s}_{j}^{\prime\Delta^{\prime}},\bar{a}_{k}^{\prime\Delta^{\prime}})+(1-\lambda)maxQ_{\theta_{j}^{\prime}}(\tilde{s}_{j}^{\prime\Delta^{\prime}},\bar{a}_{k}^{\prime\Delta^{\prime}})\right]
\end{equation}
\vspace{-0.6cm}
\end{figure*}

The above section details the training process of the network parameters of the trajectory elements $\theta_s$, $\theta_a$, $\theta_r$ and $\theta_s\prime$ in the trajectory generation model $G(\cdot)$ before deployment in the cloud. Furthermore, the parameters will be used in the denoising process to construct the complete trajectory generation model $G(\cdot)$, as shown in Fig. 6 (b). The $G(\cdot)$ model will be deployed in the cloud to generate the optimized trajectory of the target network $\Delta^\prime$. The optimization model is trained based on the generated trajectory to solve the resource allocation problem for the target network $\Delta^\prime$. It should be noted that in the expert data model of Section II, we have emphasized the overflow problem of the value range of the generated data. Therefore, BKB is integrated in $G(\cdot)$ (i.e., $G(\cdot|\mathcal{B})$) to solve the above problem. The distribution relationship $f(\cdot)$ stored in BKB $\mathcal{B}$ is used to constrain the data range of the generated trajectory so that the generated trajectory data value is within a reasonable range. Therefore, based on the trained AMLP model $\epsilon_\theta$, the constructed $G(\cdot|\mathcal{B})$ model generation process is shown in Eqs. (18), (19), (20) and (21), where $clip(\cdot)$ is a function that limits the distribution range of data generated in a single step, $\alpha^{\Delta^\prime}$ and $\beta^{\Delta^\prime}$ are the distribution ranges constraint of the target WNI $\Delta^\prime $ data values stored in BKB $\mathcal{B}$, and $\boldsymbol{z}\sim\mathcal{N}(0,{I})$. After the $G(\cdot|\mathcal{B})$ model denoises H steps, it generates a set of optimization trajectories for resource allocation optimization problems under the target WNI $\Delta^\prime$ $\tilde{\tau}^{\Delta^{\prime}} = (\tilde{s}^{\Delta^{\prime}},\tilde{a}^{\Delta^{\prime}},\tilde{r}^{\Delta^{\prime}},\tilde{s^{\prime}}^{\Delta^{\prime}})$ (i.e. $G(\cdot\mid\mathcal{B})\stackrel{\Delta^{\prime}}{\Rightarrow}\tilde{\tau}^{\Delta^{\prime}}$), which is shown in Algorithm 2.

\begin{algorithm}[!h]
\DontPrintSemicolon
  \textbf{Input:} Trained AMLP model $\epsilon_{\theta_s}$,$\epsilon_{\theta_a}$,$\epsilon_{\theta_r}$,$\epsilon_{\theta_s^{\prime}}$,WNI $\Delta^{\prime}$ of target optimization network, background knowledge base $\mathcal{B}$, Noise coefficient $\alpha_{t}$ varying with time $t$, time steps $T$.\;
    Random trajectory element sampling $g_T,p_T,\mathcal{G}_T,g^{\prime}_T{\sim}\mathcal{N}(0,I)$\;
    Extract the WNI feature $\mathcal{U}{(\Delta^{\prime})}$ of target communication network $\Delta^{\prime}$ according to Eq. (12)\;
     \For{$t=T$ to $1$}    
        { 
            Random noise sampling $\boldsymbol{z}{\sim}\mathcal{N}(0,I)$ if $t>1$;else $\boldsymbol{z}=0$\;
            Generate trajectory elements $(\tilde{g}_{t-1},\tilde{p}_{t-1},\tilde{\mathcal{G}}_{t-1},\tilde{g}_{t-1}^{\prime})$ according to Eqs. (18-21)\;
        }
    Restore trajectory elements data values $(\tilde{g}_0,\tilde{p}_0,\tilde{\mathcal{G}}_0,\tilde{g}^{\prime}_0)$ of target WNI network  $\Delta^{\prime}$ based on the mean and variance of stored in $\mathcal{B}$\;
 \textbf{Output:} Generative target WNI network optimization trajectory $(\tilde{g}_0,\tilde{p}_0,\tilde{\mathcal{G}}_0,\tilde{g}^{\prime}{}_0|\Delta^{\prime})$\;
\caption{WNI-guided trajectory generative model pipeline for power allocation optimization }
\end{algorithm}

\subsection{Generative Trajectory-based Optimization Model Construction for Resource Allocation Optimization}

We employ the generative model $G(\cdot)$ to produce optimization trajectories for the target intent network $\Delta^\prime$, which can generate sufficient set of trajectory tuples $\tilde{D}_{\Delta^\prime} = \{\tilde{\tau}_1^{\Delta^\prime}, \tilde{\tau}_2^{\Delta^\prime}, \dots, \tilde{\tau}_j^{\Delta^\prime}, \dots, \tilde{\tau}_K^{\Delta^\prime}\}$ for encoder network $E_{\text{OFF}}(\cdot)$. The encoder $E_{\text{OFF}}(\cdot)$ is deployed on the cloud, where $\tilde{D}_{\Delta^\prime}$ is applied to the encoder to establish a resource allocation optimization model for the intent network $\Delta^\prime$. The design of the encoder $E_{\text{OFF}}(\cdot)$ follows the offline DRL, which can address the extrapolation error \cite{add6,R17} associated with offline model training.

\begin{algorithm}[!h]
\DontPrintSemicolon
  Initialize Q-network $Q_{\theta_{1}}$,$Q_{\theta_{2}}$, perturbation network $\xi$, VAE model $V\!AE_{\omega}(\cdot)$, target network $Q_{\theta_{1}^{'}}$,$Q_{\theta_{2}^{'}}$\;
  \textbf{Input:} Trajectory generative model $G(\cdot|\mathcal{B})$, target communication network WNI $\Delta^\prime$, Generative numbers $L$, mini-batch size $F_2$, max perturbation $\Phi$, target network update rate $r$, minimum weighting $\lambda$.\;
     \For{$j=1$ to $L$}    
        { 
            Generate optimization trajectory of WNI target Communication network $\Delta^\prime$:  $G(\cdot|\mathcal{B})\to(\tilde{s}_{j}^{\Delta^{\prime}},\tilde{a}_{j}^{\Delta^{\prime}},\tilde{r}_{j}^{\Delta^{\prime}},\tilde{s}_{j}^{\Delta^{\prime}})=(\tilde{g}_{j}^{\Delta^{\prime}},\tilde{p}_{j}^{\Delta^{\prime}},\mathcal{G}_{j}(\Delta^{\prime}),\tilde{g}_{j}^{\prime\Delta^{\prime}})$\;
            Store generative optimization trajectory $(\tilde{g}_{j}^{\Delta^{\prime}},\tilde{p}_{j}^{\Delta^{\prime}},\mathcal{G}_{j}(\Delta^{\prime}),\tilde{g}_{j}^{\prime\Delta^{\prime}})$ in $\widetilde{D}_{\Delta^{\prime}}$\;
        }
     \For{$m=1$ to $M_{2}$}
        {
            Sample mini-batch of $F_2$ optimization trajectory from $\widetilde{D}_{\Delta^{\prime}}$\;
            Update VAE model parameter $\omega$ according to Eq. (22)\;
            Select next-time power allocation $\bar{p^{\prime}}_{j}^{\Delta_{i}}$ according to Eq. (21)\;
            Perturb each next-time power allocation: $\overline{p'}_j^{\Delta_i}=\overline{p'}_j^{\Delta_i}+\xi_\phi(\tilde{g'}_j^{\Delta'},\overline{p'}_j^{\Delta_i},\Phi)$\;
            Set value target $y$ according to Eq. (24)\;
            Update Q-network:    $\theta\leftarrow\mathrm{argmin}_{\theta} \Sigma(y-Q_{\theta}(\tilde{g}_{j}^{\Delta^{'}},\tilde{p}_{j}^{\Delta^{'}}))^{2}$\;
            Update perturb network according to Eq. (23)\;
            Update target networks: $\theta_i^{\prime}\leftarrow\tau\theta+(1-\tau)\theta_i^{\prime}$\;
            Update target perturbation network: $\phi^{\prime}\leftarrow\tau\phi+(1-\tau)\phi^{\prime}$\;
        }       
 \textbf{Output:} Resource allocation optimization policy $\pi(\tilde{g}^{\Delta^{\prime}};\omega,\phi))$ for target communication network $\Delta^{\prime}$\;
\caption{Generative trajectory-based optimization model construction for power allocation of target communication network}
\end{algorithm} 

Based on the offline DRL batch constrained deep Q-learning (BCQ) \cite{R17}, this paper performs offline model training on the trajectory tuple $\tilde{D}_{\Delta^\prime}$ generated by $G(\cdot)$. In this process, the VAE generation model $V\!AE(\cdot)$ is employed as the policy $\pi(\cdot)$ for action sampling, mitigating the extrapolation error in the generated trajectory $\tilde{s}'^{\Delta'} \rightarrow \tilde{a}'^{\Delta'}$. Additionally, to ensure the diversity of sampled actions, a perturbation model is incorporated into BCQ. Consequently, the offline generation policy, policy network update, and perturbation network update for the generated trajectory $\tilde{D}_{\Delta^\prime}$ is rewritten \cite{R17} as follows:

\setcounter{equation}{20}
\begin{equation}
\begin{aligned}\pi(\tilde{s}_{j}^{\Delta^\prime})&=\operatorname*{argmax}_{\bar{a}_{k}^{\Delta^\prime}+\xi_{\phi}(s,\bar{a}_{k}^{\Delta^\prime},\Phi)}Q_{\theta}(\tilde{s}_{j}^{\Delta^\prime},\bar{a}_{k}^{\Delta^\prime}+\xi_{\phi}(\tilde{s}_{j}^{\Delta^\prime},\bar{a}_{k}^{\Delta^\prime},\Phi))\\&\{\bar{a}_{k}^{\Delta^\prime}\sim V\!AE_{\omega}(\tilde{s}_{j}^{\Delta^\prime})\}_{k=1}^{n}\end{aligned}
\end{equation}
\begin{equation}
\omega\leftarrow\mathrm{argmin}_\omega\Sigma(\tilde{a}_j^{\Delta^\prime}-\bar{a}_k^{\Delta^\prime})^2+D_\mathrm{KL}(\mathcal{N}(\mu,\sigma)||\mathcal{N}(0,1))
\end{equation}
\begin{equation}
    \phi\leftarrow\mathrm{argmax}_\phi\Sigma Q_{\theta}\left(\widetilde{s}_j^{\Delta^\prime},\overline{a}_k^{\Delta^\prime}+\xi_\phi\left(\widetilde{s}_j^{\Delta^\prime},\overline{a}_k^{\Delta^\prime},\Phi\right)\right)
\end{equation}
where $\phi$ denotes the model parameter of the perturbation network $\xi$, $D_{KL}$ denotes the KL divergence, $\mathcal{N}(\mu,\sigma)$ denotes  the distribution calculated by $V\!AE (\cdot)$ from the generated trajectory elements $(\tilde{s}^{\Delta^\prime}, \tilde{a}^{\Delta^\prime})$, $\boldsymbol{\phi}$ denotes the maximum perturbation value, and $\omega$ denotes the VAE model parameter. In addition, the critic network uses soft-clipped double Q-learning, so the target value is expressed as in Eq. (24).

\begin{figure*}[]
  \begin{center}
   \subfigure[Channel gain ($s_{\mathbf{\Delta}}$)]{
    \scalebox{0.5}[0.5]{\includegraphics{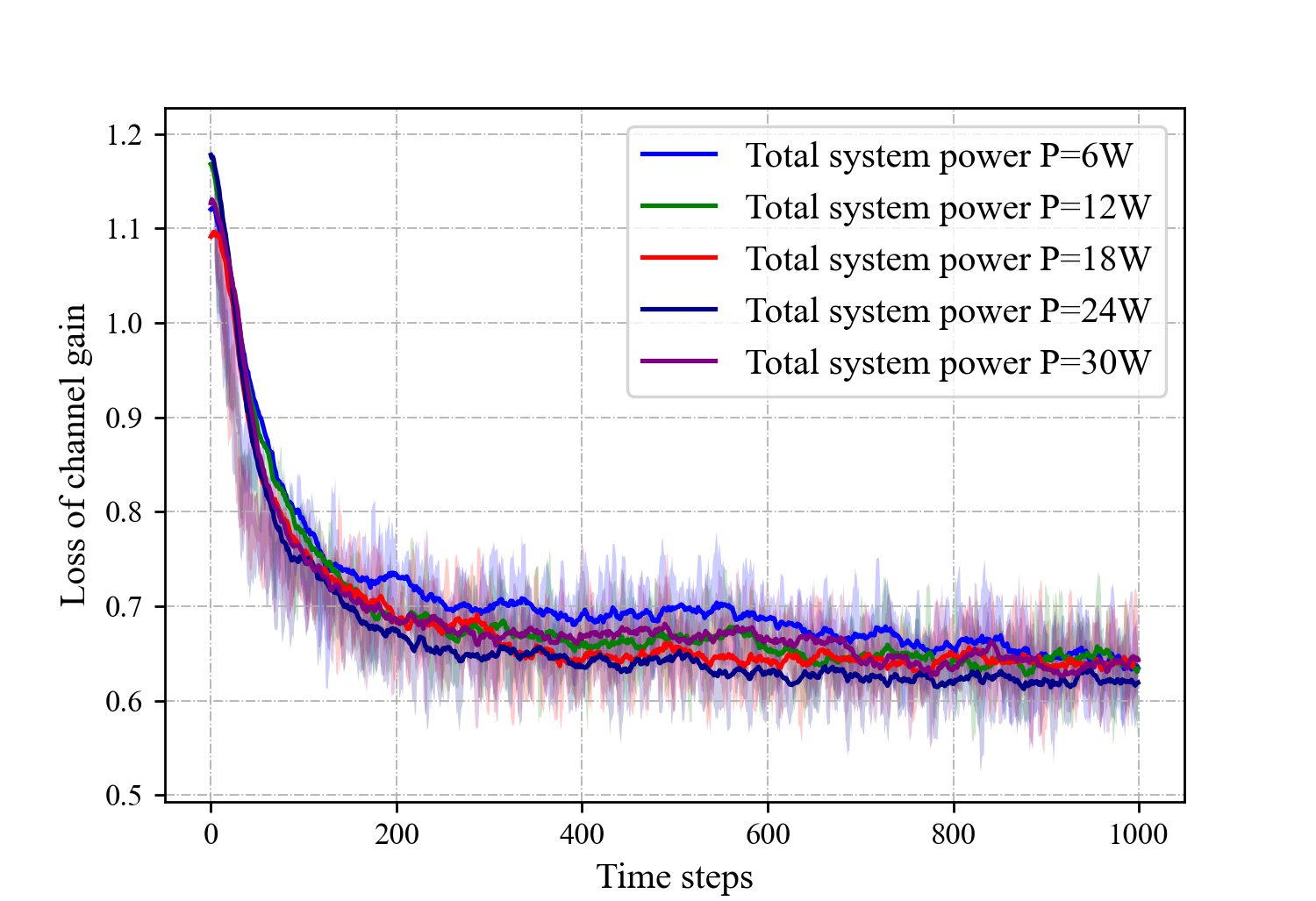}}  
    \label{fig:2}
    }
     \subfigure[Power allocation ($a_{\mathbf{\Delta}}$)]{
    \scalebox{0.5}[0.5]{\includegraphics{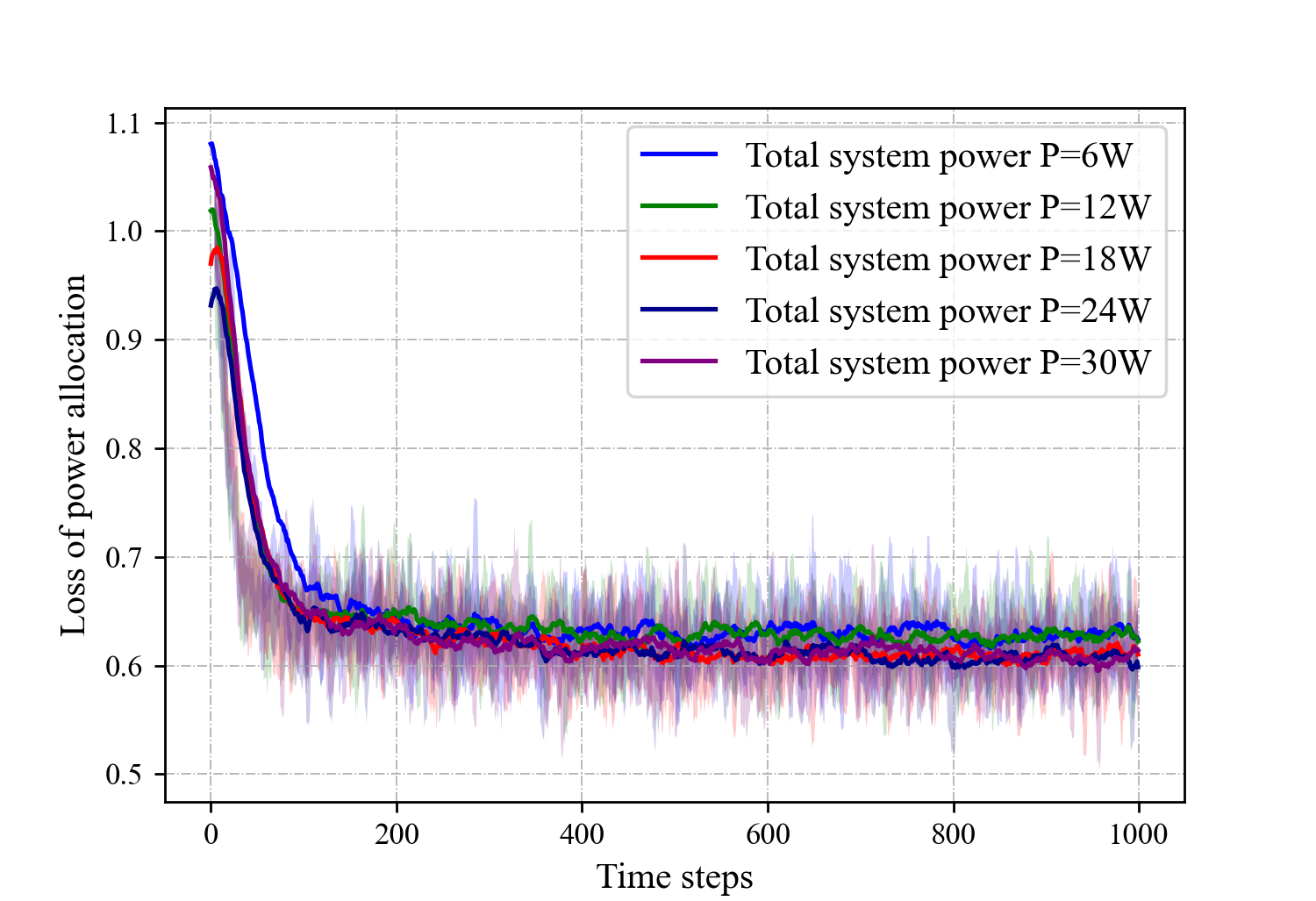}}
    \label{fig:3}
    }   
     \subfigure[Spectral efficiency ($r_{\mathbf{\Delta}}$)]{
    \scalebox{0.5}[0.5]{\includegraphics{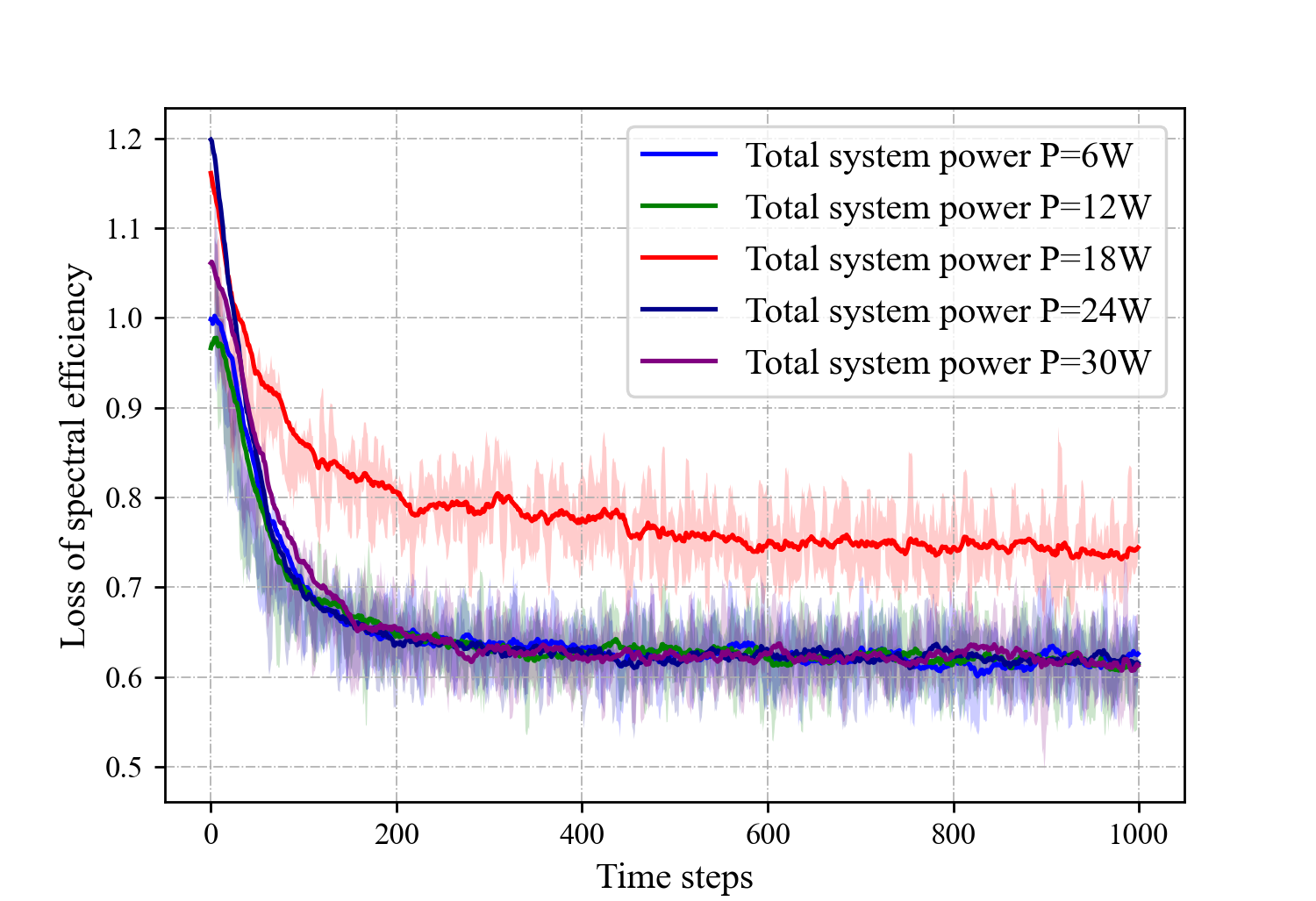}}  
    \label{fig:2}
    }
     \subfigure[Next time channel gain ($s'_{\mathbf{\Delta}}$)]{
    \scalebox{0.5}[0.5]{\includegraphics{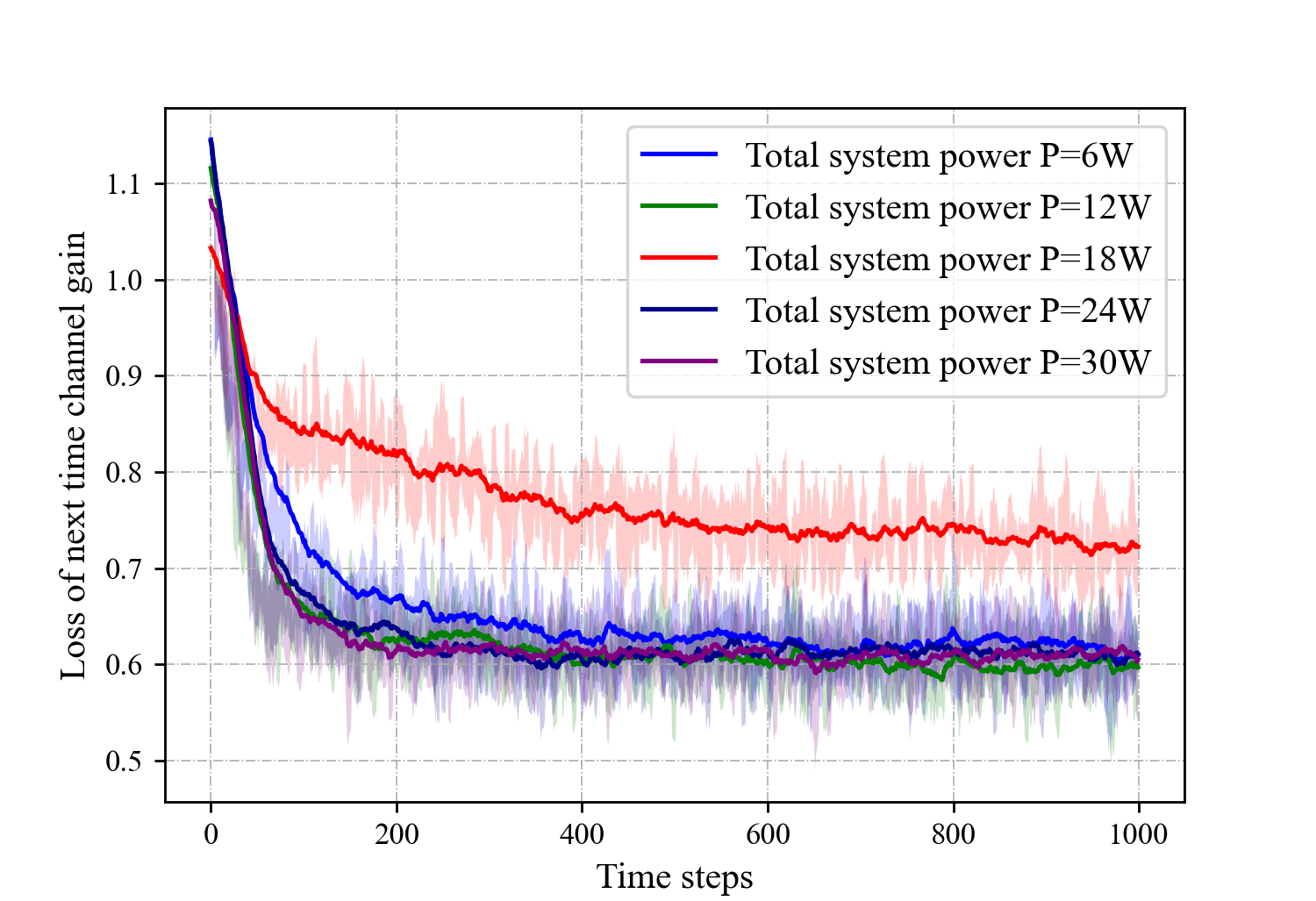}}
    \label{fig:3}
    } 
    \caption{The training loss of WNI-guided trajectory generative model for different total system communication transmission power.}
  \end{center}
  \vspace{-0.6cm}
\end{figure*}

The optimization model construction process of the $E_{OFF}(\cdot)$ encoder is shown in Algorithm 3. The $E_{OFF}(\cdot)$ encoder constructs the optimization policy $\pi(\tilde{s}^{\Delta^{\prime}};\omega,\phi))$ through the intent trajectory generated by $G(\cdot)$, which can be directly deployed in the intent $\Delta^\prime$ network to perform fine tuning resource allocation optimization operations. The proposed scheme avoids the limitations of traditional online DRL methods that require real-time interaction with the actual network and traditional offline DRL without training datasets.

\setcounter{equation}{24}

\subsection{Complexity Analysis}
The computational complexity of the proposed optimization scheme in this paper can be mainly divided into the following two parts:

\textbf{Training and Generation of WNI-Guided Optimization Trajectory Generation Model}: The proposed model training and generation complexity is $O(4(M_{1} F_{1}+L b)(T|\theta|+|\delta|))$.
During the entire training phase, which involves $M_{1}$ training steps, four AMLP networks are used. Each training sample contains $F_{1}$ optimization trajectories. For each sampled optimization trajectory, the training process includes $T$ steps of DM-based adding noise. In addition, the number of WNI encoder parameters is $|\delta|$, and the computational complexity of each step for each AMLP network is $O(|\theta|)$. Therefore, the computational complexity of the generative model training process is $O(4M_{1}F_{1}(T|\theta|+|\delta|)$. For the generation phase, the process involves producing $b$ target intent categories and generating $L$ trajectories for each category. As a result, the computational complexity of the generation phase is $O\left(4Lb(T|\theta| + |\delta|)\right)$.

\textbf{Generative Optimization Trajectory-based Offline DRL Policy Learning}: Offline DRL optimization policy training has a computational complexity of $O({F}_{2}M_{2}(|\omega|+|\phi|+|\theta_{Q}|)+M_{2}|(|\theta_{Q}|+|\phi|))$. The training process consists of $M_{2}$ rounds of offline policy updates, with $F_{2}$ generated trajectories sampled in each round.  The policy network parameters include the VAE generation network parameters and perturbation network parameters $|\omega|$ and $|\phi|$, respectively. Additionally, the Q-function network contains parameters represented by $|\theta_{Q}|$. Thus, the policy network computational complexity is $O(F_{2}M_{2}(|\omega|+|\phi|))$, the Q-function network computational complexity is $O(F_{2}M_{2}|\theta_{Q}|)$, and the target Q-function network and target perturbation network computational complexity is $O(M_{2}(|\theta_{Q}|+|\phi|)$.

Therefore, the computational complexity of the proposed optimization algorithm in this paper is $O(4(M_{1} F_{1}+L b)(T|\theta|+|\delta|)+{F}_{2}M_{2}(|\omega|+|\phi|+|\theta_{Q}|)+M_{2}(|\theta_{Q}|+|\phi|))$.

\section{Simulation Results}

This section fully considers the model construction paradigm of Section II, mainly including the collection and preprocessing of expert datasets, as well as the construction of the generation system and performance simulation testing.

\begin{table}[h]
\centering
\caption{Simulation parameters}%
\begin{tabular}{l l l l}
\hline %
\textbf{Parameters for Communication System} & {\textbf{Values}} \\
\hline%
The number of intelligence devices & $16$\\
\hline
Total system transmission power (W)	& $\{6,12,18,24,30\}$\\
\hline
The distribution range of channel gain  (dB)	& ${{[0,50)}}$\\
\hline
Number of completely independent \\ channel gain levels & 5	\\
\hline
Number of completely independent \\ WNI categories 	& 5	\\
\hline
Total number of expert optimization trajectory	& 1000000	\\
\hline %
\textbf{Parameters for AMLP-based DM} & {\textbf{Values}} \\
\hline%
Actor NN Learning Rate & 0.0002\\
\hline
Critic NN Learning Rate & 0.0001\\
\hline
Soft Update Parameter & 0.005\\
\hline
Mini-batch Size & 64\\
\hline
Hidden layer dimension & 64\\
\hline
Number of attention heads & 4\\
\hline
Attention head dimension & 8\\
\hline
WNI feature dimension & 16\\
\hline
Time embedding dimension & 16\\
\hline
Time steps for DM & 5\\
\hline
The range of noise coefficient & $[0.0001,0.02]$\\
\hline
Number of layers & 4\\
\hline %
\textbf{Parameters for offline DRL} & {\textbf{Values}} \\
\hline%
Discount factor & 0.1\\
\hline
Mini-batch size	& 100\\
\hline
Weighting for clipped double Q-learning	& 0.75\\
\hline
Soft update parameter & 0.1	\\
\hline
Max perturbation hyper-parameter & 0.05	\\
\hline
Hidden layer dimension	& 32	\\
\hline
\end{tabular}
\end{table}

\subsection{Experimental Setup}
 Without loss of generality, to ensure the relevance of the intention features and the distribution of trajectory data, the intention features of this paper are distinguished and marked as 5 different intention feature vectors, which represent ``low channel gain scenario'', ``lower channel gain scenario'', ``medium channel gain scenario'', ``high channel gain scenario'' and ``very high channel gain scenario''. For illustration purposes, we provide the distance-dependent path loss model \cite{add7} of the channel is expressed as:
 \begin{equation}
     L(d) = C_0 \left( \frac{d}{D_0} \right)^{-\gamma}
 \end{equation}
where $C_0$ denotes the path loss at the reference distance $D_0 =1m$, $d$ represents the individual link distance, and $\gamma$ is the path loss exponent. To align with the optimization problem in Eq. (8), the channel gain is given by:
 \begin{equation}
 g = 10^{\frac{G_t + G_r - L(d)}{10}}
\end{equation}
where $G_t$ and $G_r$ represent the transmit antenna gain and receive antenna gain, respectively. We hypothesize that the five intent scenarios under consideration correspond to distinct parameter configurations $(G_t, G_r, \gamma, d, C_0, D_0)$, which collectively affect the range of the channel gain $g$ distribution for each scenario. For clarity in discussion and experimental validation, we assume that the different intent scenarios 1-5 correspond to channel gain values falling within the ranges (0, 10), [10, 20), [20, 30), [30, 40), and [40, 50), respectively.
 
Additionally, to ensure the effectiveness of the source of the expert dataset, this paper uses the water-filling algorithm power allocation collection paradigm provided by \cite{R9} to realize the collection of expert datasets. A total of 1,000,000 power allocation expert trajectory data samples were collected within the above five design WNI and processed by $\mathcal{N}{\sim}(0,1)$. The intention distribution relationship and the mean-variance of the expert data were stored in BKB and finally integrated into the $G(\cdot)$ model, which is used to guide the generation of the optimization trajectory. There may be intersections in the numerical distribution, but this does not affect the practical application of the system paradigm in this paper. Additionally, considering the low-dimensional characteristic of the wireless communication optimization problem\cite{R5,R7,R18}, the number of hidden layers is limited to 3-5 to avoid unfavorable effects on trajectory feature learning, such as the inability to effectively capture the essential characteristics of the optimization trajectories, gradient vanishing or explosion, etc. The core simulation system parameters are summarized in Table I.

\subsection{Simulation Result}

 To verify the convergence and effectiveness of the intent-guided power allocation optimization trajectory generation model developed in this paper, we train the expert dataset, which integrates five independent distribution intents. In the experiment, five total communication transmission power configurations (i.e. 6W, 12W, 18W, 24W, and 30W) were selected to address the power allocation optimization problem under varying system total power conditions. Fig. 7 illustrates the training loss of the AMLP network within the generation model, which is based on channel gain, power allocation, spectral efficiency, and the next-time channel gain of the optimized trajectory tuple under the different total power configurations. A lower loss value indicates a smaller discrepancy between the noise predicted by the AMLP model and the actual value, directly impacting the model's effectiveness. Under different total power constraints, the AMLP network converged within 400 iterations, reducing the loss to approximately 0.6.

Fig. 8 illustrates the scatter plot of the trajectory data generated after different wireless network intents are input into the trajectory generation model and the accuracy of the trajectory distribution. Each color in the scatter plot represents a different input intent category, and each intent category contains 1600 sampled generated optimized trajectory points. Each step of the DM denoising process in the generative model is constrained by the functional relationship of the BKB. This ensures that the generated results (i.e., including channel gain, power allocation, spectral efficiency, and the subsequent channel gain trajectory tuple) remain within the normal distribution range corresponding to the intended values. Ultimately, this approach guides the restoration of the optimized trajectory values. The experimental results, illustrated in the scatter plot, demonstrate that the generative model, which integrates the BKB relationship, successfully positions all generated trajectory points within the defined distribution range. Furthermore, the accuracy of the distribution for all trajectory points within this range is 100\%, which addresses the data boundary overflow issue encountered in the generative model for wireless communication optimization trajectory generation. This ensures the effectiveness of the generative DRL optimization trajectories.


 \begin{figure}
   \begin{center}
    \scalebox{0.26}[0.26]{\includegraphics{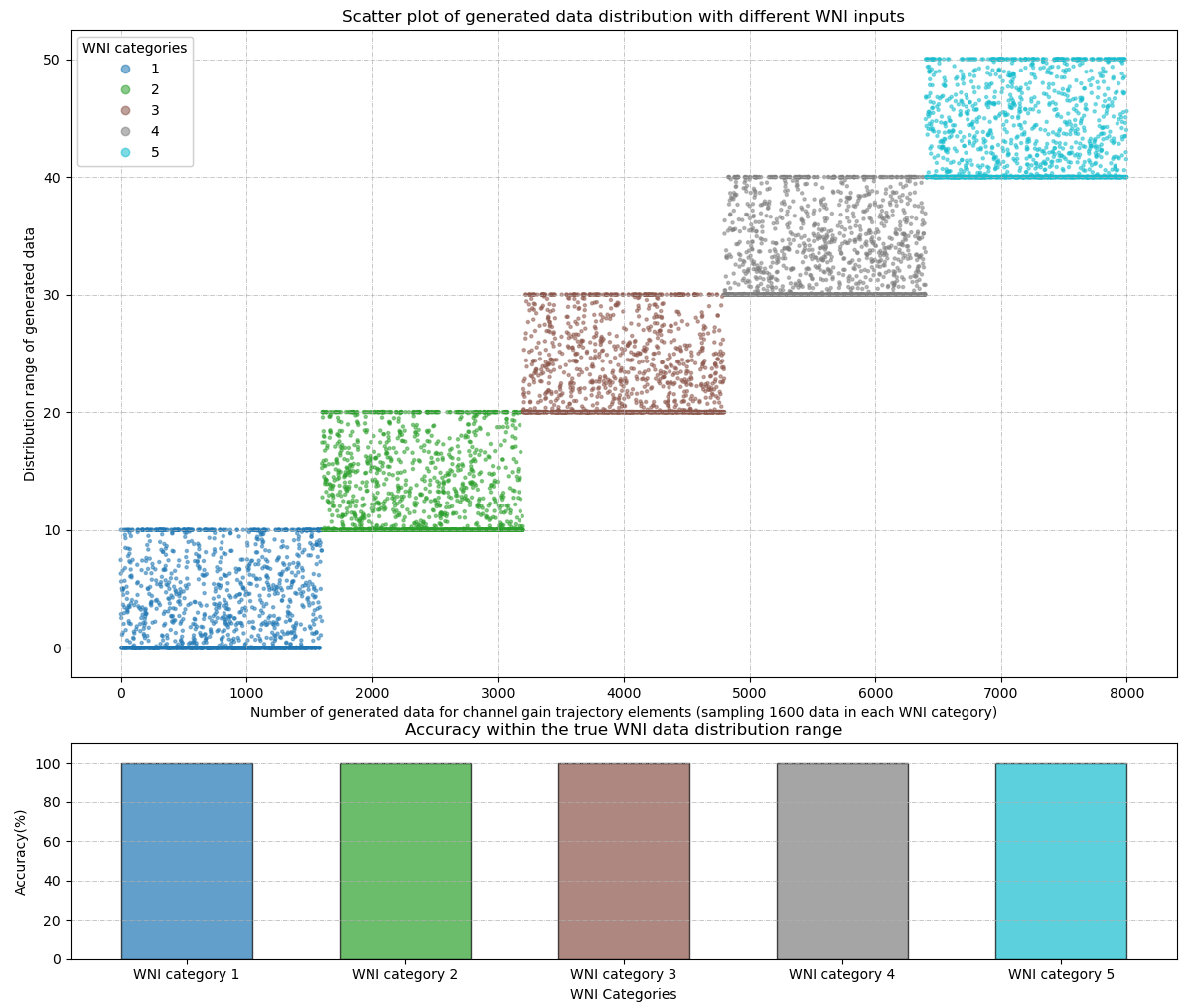}}
    \caption{Generative channel gain trajectory scatter plots and distribution accuracy by generative model $G (\cdot)$ for different WNI inputs.}
    \label{fig:2}
  \end{center}
  \vspace{-0.6cm}
\end{figure}

\begin{figure}
   \begin{center}
    \scalebox{0.65}[0.65]{\includegraphics{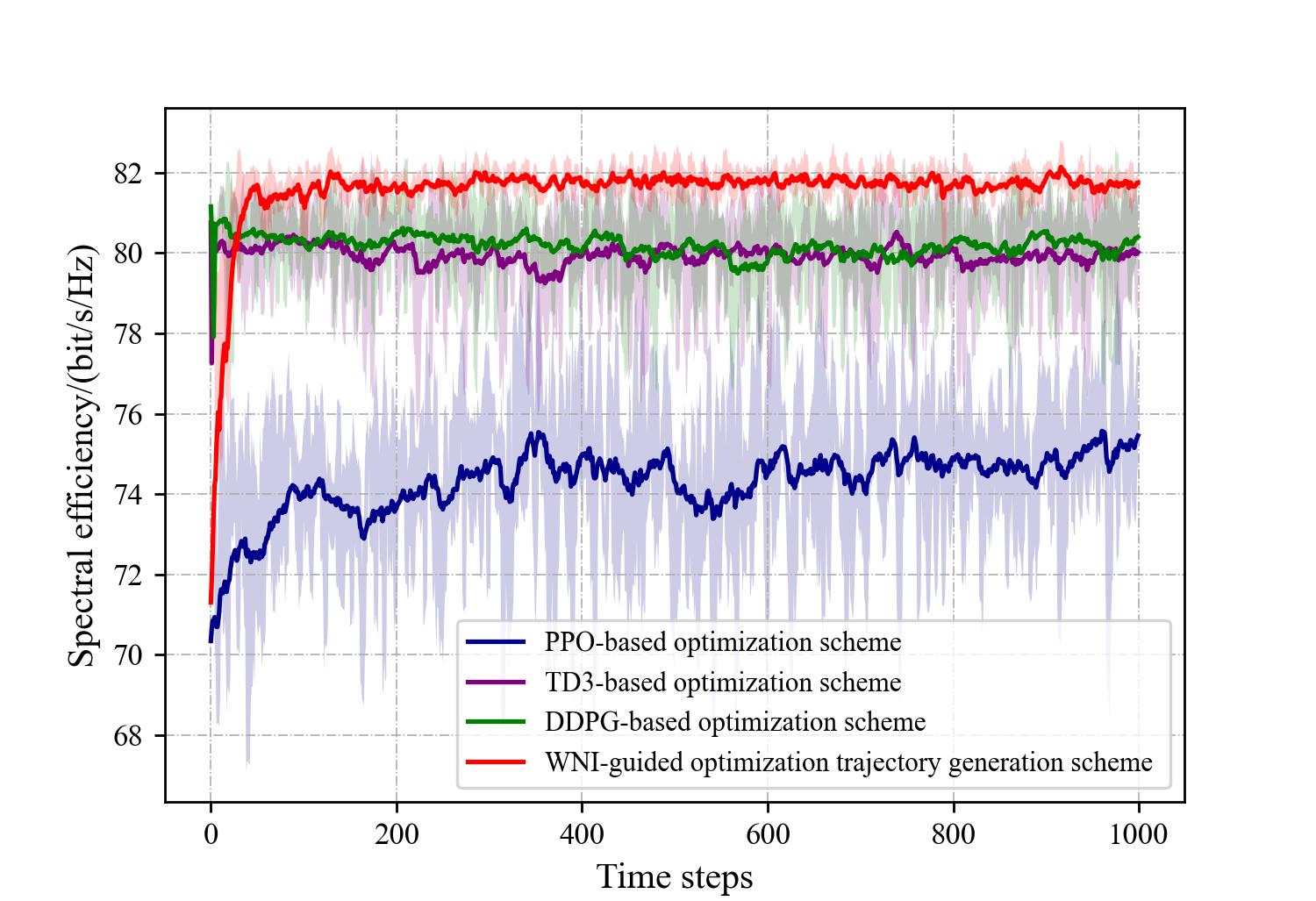}}
    \caption{An example of spectral efficiency versus time steps under different schemes for one of the WNIs.}
    \label{fig:2}
  \end{center}
  \vspace{-0.6cm}
\end{figure}

\begin{figure*}[]
  \begin{center}
   \subfigure[wireless network intention category 1]{
    \scalebox{0.5}[0.5]{\includegraphics{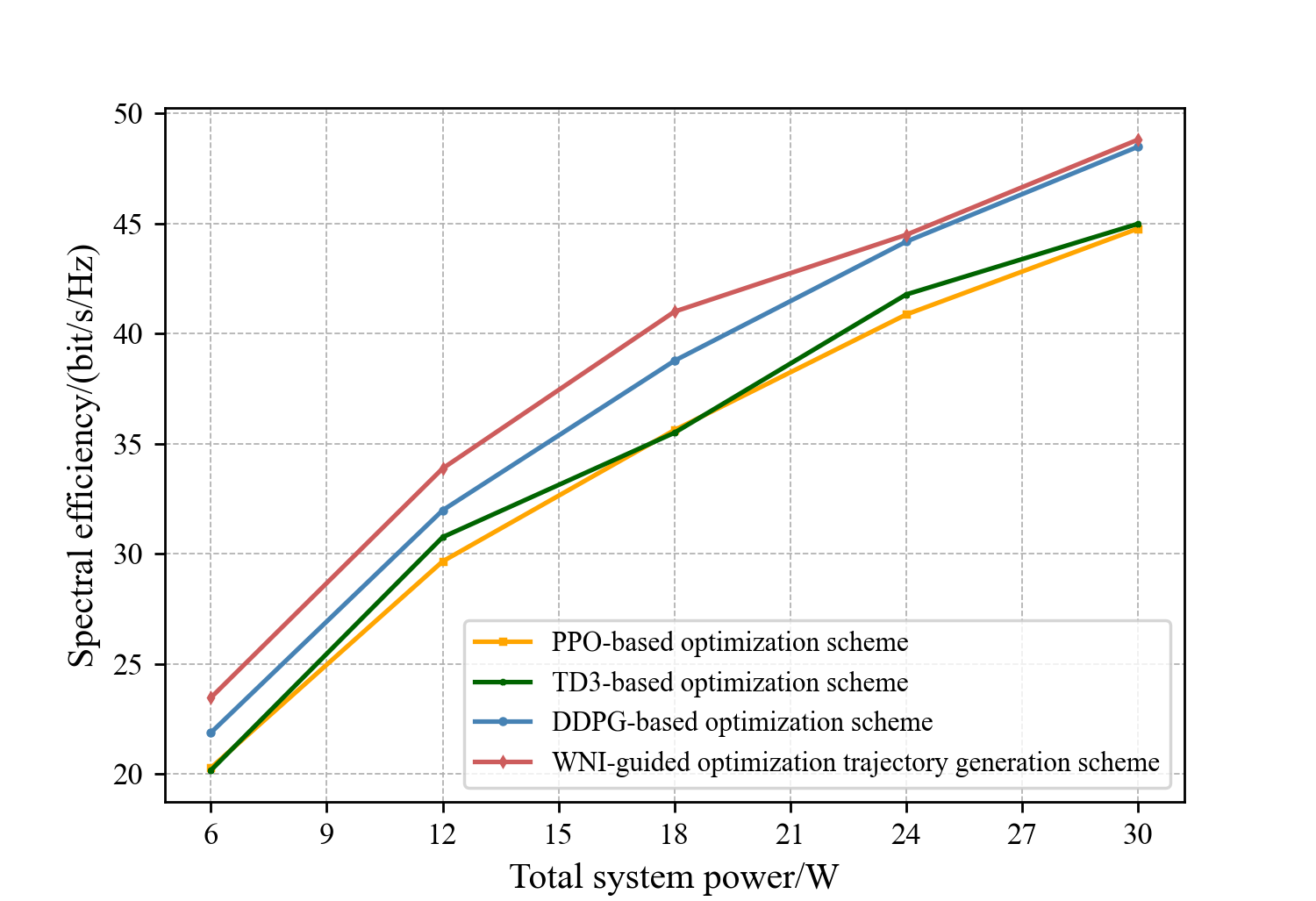}}  
    \label{fig:2}
    }
     \subfigure[wireless network intention category 2]{
    \scalebox{0.5}[0.5]{\includegraphics{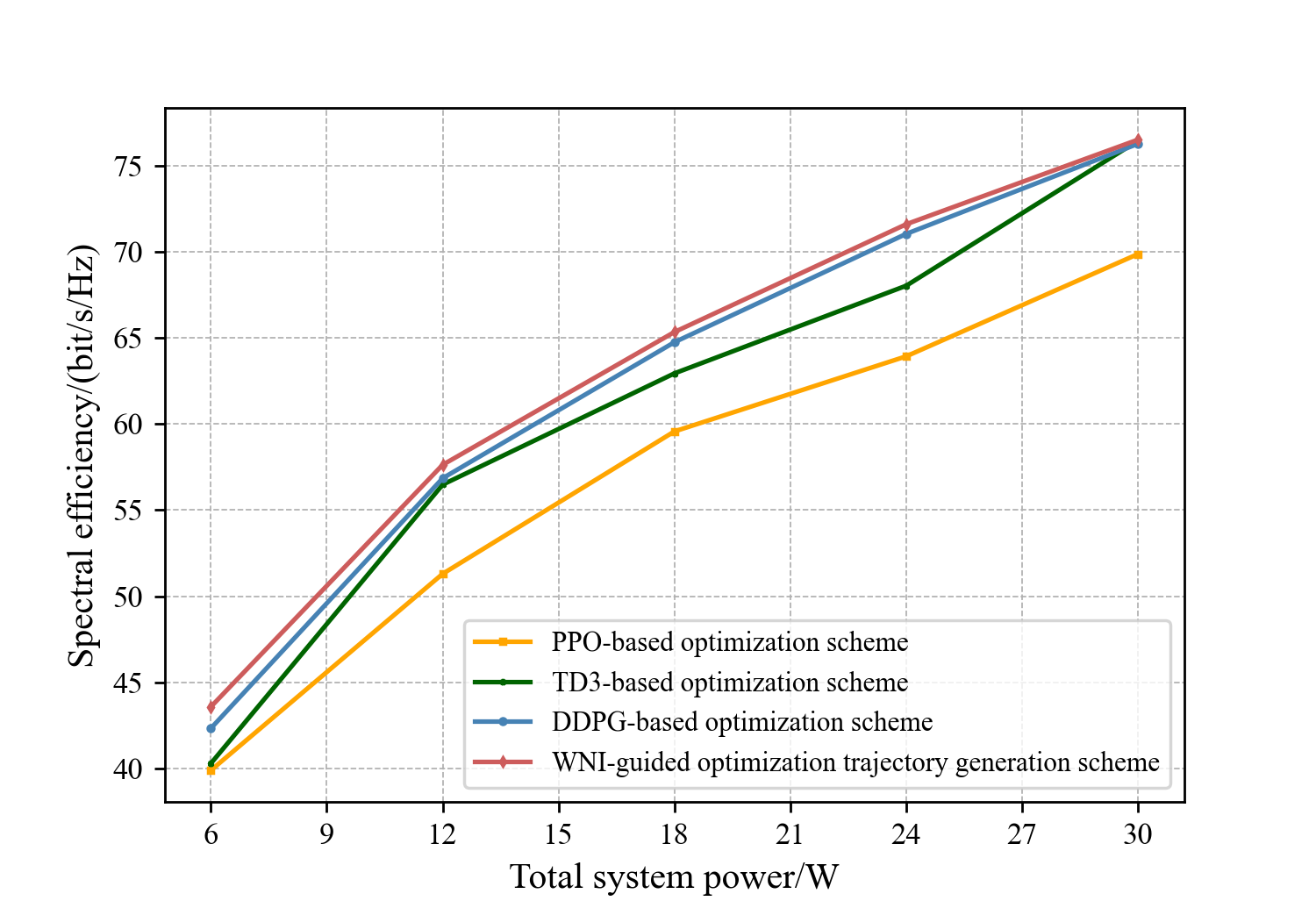}}
    \label{fig:3}
    }   
     \subfigure[wireless network intention category 3]{
    \scalebox{0.5}[0.5]{\includegraphics{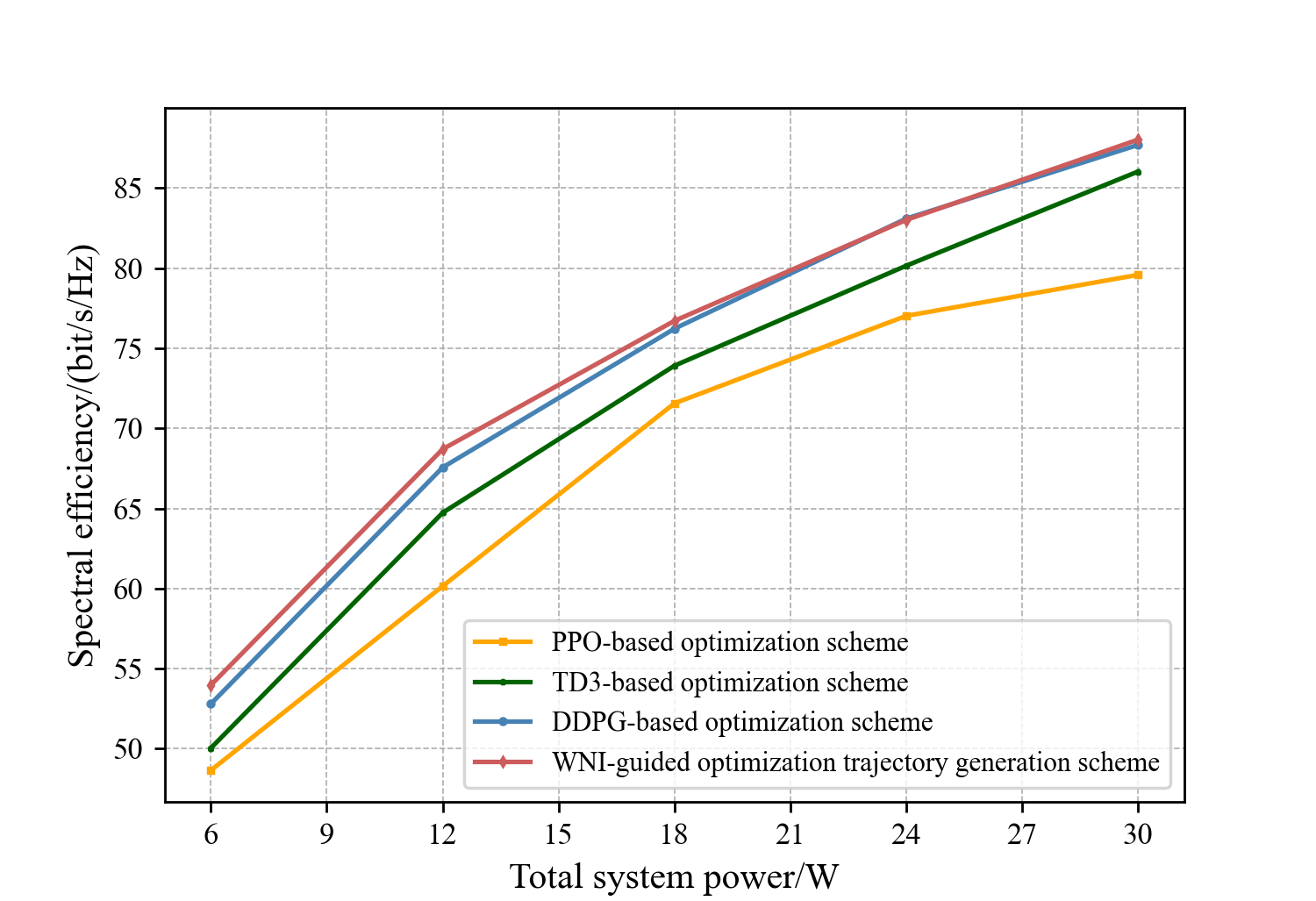}}  
    \label{fig:2}
    }
     \subfigure[wireless network intention category 4]{
    \scalebox{0.5}[0.5]{\includegraphics{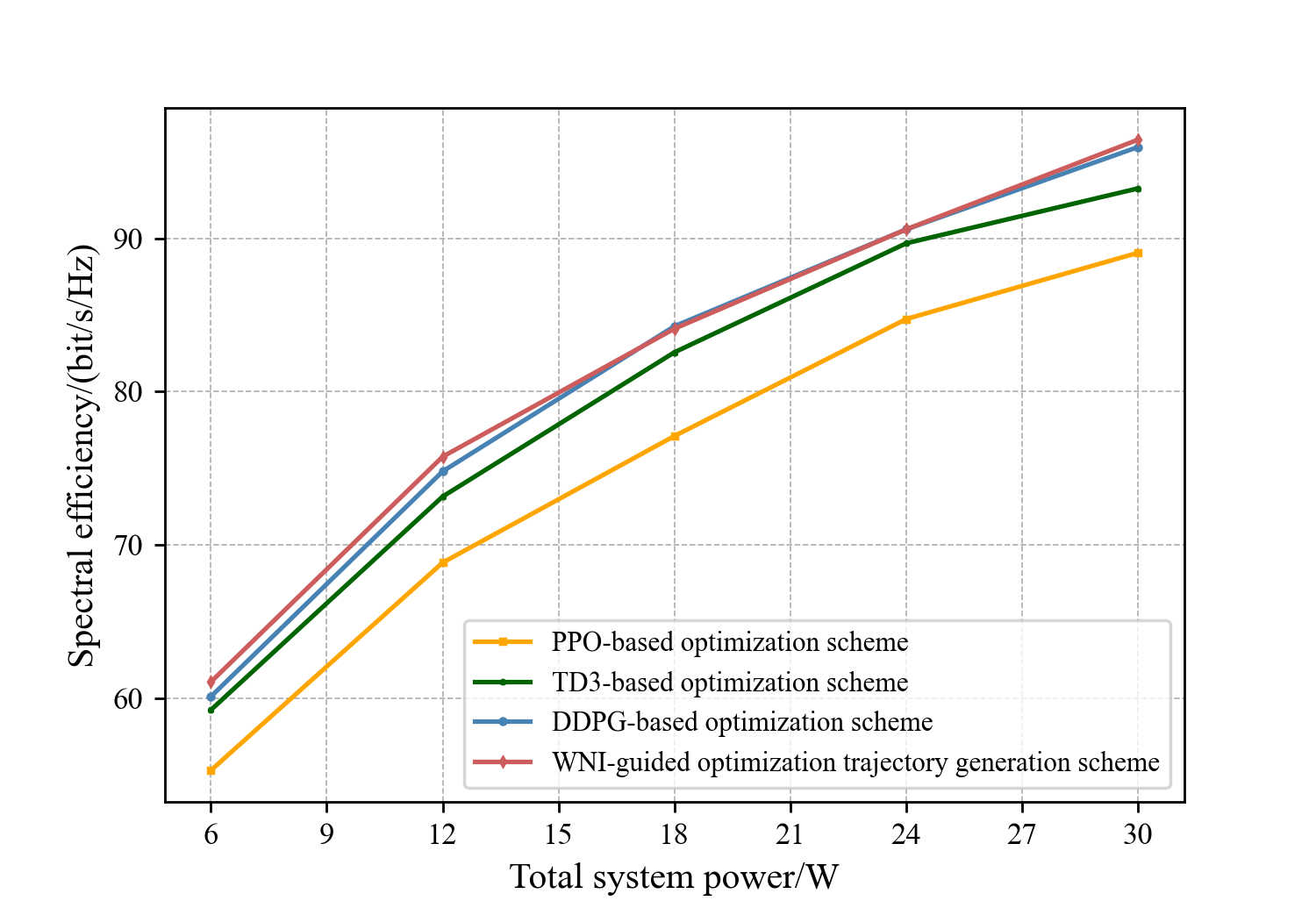}}
    \label{fig:3}
    } 
     \subfigure[wireless network intention category 5]{
    \scalebox{0.5}[0.5]{\includegraphics{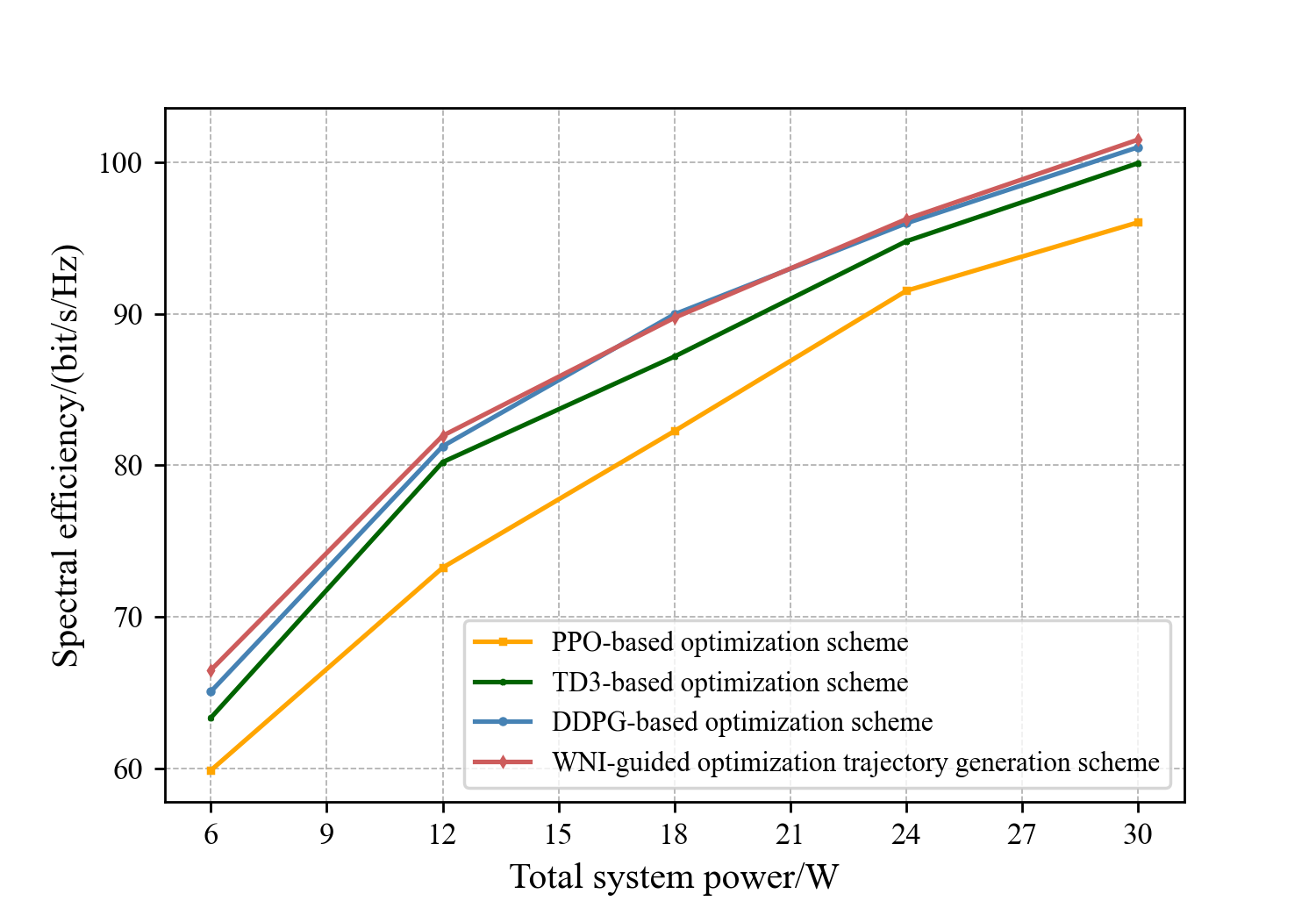}}
    \label{fig:3}
    } 
    \caption{Spectrum efficiency performance comparison versus different total system power in various WNI category scenarios under different optimization schemes.}
  \end{center}
  \vspace{-0.6cm}
\end{figure*}

Fig. 9 illustrates the performance of the proposed WNI-guided optimization trajectory generation scheme with the DDPG-based, TD3-based, and PPO-based optimization scheme over time steps. Among these schemes, the DDPG-based and TD3-based optimization scheme exhibit relatively stable performance within 10 steps in the online optimization process, due to the trajectory values remaining within a consistent numerical range. In contrast, the PPO-based optimization scheme performs less effectively. Moreover, the proposed GDM-aided fine-tuning optimization scheme stabilizes at around 100 steps. From the perspective of training stability, the GDM-aided optimization scheme demonstrates greater stability compared to the online DRL methods in the same channel environment. Additionally, the spectral efficiency achieved by the GDM-aided optimization scheme stabilizes at approximately 82 bits/s/Hz, outperforming the DDPG-based and TD3-based schemes, which stabilize around 80 bits/s/Hz, and the PPO-based scheme, which stabilizes around 75 bits/s/Hz. This advantage can be attributed to the generated trajectory, which provides the optimization model with expert trajectory data distribution in the intention-based communication scenario before model deployment. The offline optimization scheme, based on intent-generated trajectories, learns the communication scenario’s optimization policy in advance, enabling it to better adapt to channel variations and maintain more consistent performance compared to the online DRL solutions. Similarly, the GDM-aided optimization scheme, supported by the generated trajectory dataset, is capable of finding the optimal solution for the power allocation problem.

To thoroughly evaluate the effectiveness of the WNI-guided trajectory generation optimization system proposed in this paper, we conducted detailed simulations under five independent intents and five total power settings, employing different AI-based optimization methods. The results of these simulations are illustrated in Fig. 10. Specifically, Figs. 10(a)–10(e) correspond to the performance comparison results under five WNI features: low, lower, medium, higher, and high channel gain communication scenarios. Among the tested methods, the DDPG-based, TD3-based, and PPO-based optimization schemes enable real-time online decision-making, whereas the GDM-aided optimization scheme proposed in this paper inputs the intent features into the WNI-guided trajectory generation model in advance to generate the optimized trajectory corresponding to the intent, and completes the pre-training of the optimization policy. The results shown in Fig. 10 clearly demonstrate the overall performance gains achieved by the WNI-guided trajectory generation model. Across all simulations, the GDM-aided optimization scheme consistently exhibits the highest spectrum efficiency. Notably, the performance gain is more significant when the total allocated power is set to 6W and 12W. This difference is attributed to the full power allocation strategy adopted in this paper\cite{R9}, which provides greater fault tolerance at lower power levels (6W and 12W) compared to higher power levels (18W, 24W, and 30W) under varying channel gain conditions.

Furthermore, the improvement in spectrum efficiency is largely due to the optimization experience provided by the WNI-guided trajectory generation model prior to deployment. This pre-trained experience enables the AI model to better adapt to dynamic environmental conditions and make more effective decisions in real time.

\section{Conclusion and Future Work}

This paper proposes a novel WNI-guided trajectory generation communication system, which can generate optimized trajectories under the target intent communication network, and the trajectories can help pre-train the optimization strategy. 
In practical applications of AI-driven wireless resource allocation optimization, users can employ this system to obtain optimized trajectories for the target network in advance, using intent-guided pre-training to enhance optimization strategies. This significantly reduces the constraints of traditional AI optimization methods on real-time interactive decision-making and allows for the generation of trajectory data across various target scenarios, facilitating more efficient model learning. Simulation results demonstrate that the optimized trajectories generated by this system fall within the actual intent data distribution, and the optimization model pre-trained with these trajectories exhibits superior performance compared to the online DRL-based optimization scheme.

The current system addresses a limited range of target scenarios, which suggests that future research should explore broader optimization frameworks. For instance, leveraging large language models (LLMs) \cite{b5} to integrate network design intents could provide new avenues for solving wireless network optimization challenges. Additionally, the scope of optimization model design is not confined to resource allocation, which can be extended to a variety of emerging problems such as UAV trajectory optimization, Wi-Fi standardization, and channel coding. These extensions could further enhance the efficiency and adaptability of next-generation wireless communication systems.







\bibliographystyle{IEEEtran}
\bibliography{reference}

\vspace{12pt}

\end{document}